\documentclass[lettersize,journal]{IEEEtran}
\IEEEoverridecommandlockouts
\usepackage[ruled,vlined]{algorithm2e}
\usepackage[T1]{fontenc} 
\usepackage{cite}
\usepackage{amsmath,amssymb,amsfonts}
\usepackage{graphicx}
\usepackage{xcolor}
\usepackage{array,color}
\usepackage[caption=false,font=normalsize,labelfont=sf,textfont=sf]{subfig}
\usepackage{textcomp}
\usepackage{stfloats}
\usepackage{url}
\usepackage{verbatim}
\usepackage{bm}
\usepackage{mathrsfs}
\usepackage[utf8]{inputenc}
\usepackage[english]{babel}
\usepackage[colorlinks]{hyperref}
\hypersetup{
    citecolor=blue,  
    linkcolor=blue, 
    urlcolor=cyan  
}
\begin{document}

\title{Fast Time-Varying mmWave MIMO Channel Estimation and Reconstruction: An Efficient Rank-Aware Matrix Completion Method\\}
\author{Tianyu Jiang, \IEEEmembership{Student Member, IEEE}, Yan Yang, \IEEEmembership{Senior Member, IEEE}, Hongjin Liu, Runyu Han, \\Bo Ai, \IEEEmembership{Fellow, IEEE}, Mohsen Guizani, \IEEEmembership{Fellow, IEEE}

\thanks{Part of this work has been accepted by the IEEE Global Communications Conference (GLOBECOM), Taipei, Taiwan, 2025 \cite{b1}.  This work was funded by the National Nature Science Foundation of China under Grants 62271041 and 62221001, and in part by the Hongguoyuan Science and Technology Project under Grant WX-2025-0803.\\\textit{(Corresponding author: Yan Yang.)}}
\thanks{Tianyu Jiang, Yan Yang, Runyu Han and Bo Ai are with the School of Electronic and Information Engineering, Beijing Jiaotong University, Beijing 100044, China (e-mail:jiangty@bjtu.edu.cn, yyang@bjtu.edu.cn, lhjbuaa@163.com, runyuhan@bjtu.edu.cn, boai@bjtu.edu.cn).\\
	\text{ }\text{ }\text{ }Hongjin Liu is with Beijing SunWise Information Technology Ltd., Beijing 100190, China (e-mail: lhjbuaa@163.com).\\
	\text{ }\text{ }\text{ }Mohsen Guizani is with the Department of Machine Learning, Mohamed bin Zayed University of Artificial Intelligence, Abu Dhabi, United Arab Emirates (e-mail: mguizani@ieee.org).}}
\maketitle
\begin{abstract}
We address the problem of fast time-varying channel estimation in millimeter-wave (mmWave) MIMO systems with imperfect channel state information (CSI) and facilitate efficient channel reconstruction. Specifically, leveraging the low-rank and sparse characteristics of the mmWave channel matrix, a two-phase rank-aware compressed sensing framework is proposed for efficient channel estimation and reconstruction. In the first phase, a robust rank-one matrix completion (R1MC) algorithm is used to reconstruct part of the observed channel matrix through low-rank matrix completion (LRMC). To address abrupt rank changes caused by user mobility, a discrete-time autoregressive (AR) model is established that leverages temporal rank correlations across consecutive time instances to enable adaptive observation matrix completion, thereby improving estimation accuracy under dynamic conditions. In the second phase, a rank-aware block orthogonal matching pursuit (RA-BOMP) algorithm is developed for sparse channel recovery with low computational complexity. Furthermore, a rank-aware measurement matrix design is introduced to improve angle estimation accuracy. Simulation results demonstrate that, compared with existing benchmark algorithms, the proposed approach achieves superior channel estimation performance while significantly reducing computational complexity and training overhead.
	
\end{abstract}
\section{Introduction}

\IEEEPARstart{M}{illimeter}-wave (mmWave) communication, which exploits the abundant spectrum resources in the 30\textasciitilde 300 GHz band, has emerged as a key enabler for meeting the ever-increasing capacity demands of future wireless networks \cite{b2}. In practical deployments, however, mmWave signals experience significantly greater path loss than sub-6 GHz counterparts. To overcome this limitation, beamforming techniques, particularly hybrid analog–digital (HAD) beamforming architectures integrated with massive multiple-input multiple-output (MIMO) configurations, have been recognized as effective solutions for compensating propagation loss by forming high-gain directional beams \cite{b3}. Consequently, accurate channel estimation is indispensable for achieving precise beam alignment between transmitters and receivers. Nevertheless, in massive mmWave MIMO systems, the estimation process involves high-dimensional channel matrices due to the deployment of large-scale antenna arrays, thereby significantly increasing computational complexity. Furthermore, obtaining accurate and timely channel state information (CSI) in mobile mmWave networks remains challenging due to rapid channel variations \cite{b4}.

MmWave channels typically exhibit a limited number of dominant propagation paths due to high propagation and penetration losses, resulting in inherently sparse and low-rank channel matrices \cite{b5,b6}. In narrowband systems, orthogonal matching pursuit (OMP) algorithms are widely employed to exploit angular-domain sparsity \cite{b7,b8}. Meanwhile, several spectral estimation and low-rank matrix completion (LRMC) techniques have been proposed \cite{b9}. In wideband systems such as orthogonal frequency-division multiplexing (OFDM), compressed sensing (CS) techniques have been developed to exploit joint sparsity across the time and frequency domains \cite{b10}. For instance, time-domain estimators leverage joint sparsity across angular and delay domains \cite{b11,b12}, whereas frequency-domain approaches primarily utilize angular sparsity \cite{b13,b14}. Although time-domain estimation generally provides higher accuracy, it often incurs significant computational overhead. To alleviate discretization errors in CS-based methods, low-rank tensor decomposition has been introduced for wideband frequency-selective channels \cite{b15}. However, this technique has primarily been evaluated under relatively high signal-to-noise ratio (SNR) conditions, which are rarely achievable in practical mmWave scenarios. Beyond sparse scattering characteristics, mmWave channels also exhibit angular dispersion across angles of arrival (AoAs), angles of departure (AoDs), and elevation domains, leading to structured sparsity and low-rank features that can be further exploited to enhance estimation performance \cite{b16}. This jointly sparse and low-rank structure, as investigated in \cite{b17,b18}, is leveraged by a two-stage CS framework that first performs LRMC followed by sparse recovery to reconstruct the mmWave channel. Although this approach achieves reduced sample complexity compared with conventional CS-based time-domain estimation, its effectiveness under fast time-varying channels remains unexplored.

To address the challenges associated with time-varying mmWave channel estimation, e.g., dynamic multipath sparsity, evolving low-rank channel structures, and high-dimensional optimization, a variety of adaptive techniques have been proposed. For instance, sparse Bayesian learning (SBL) models have been employed to track the emergence and disappearance of propagation paths. Although these methods offer adaptivity, their computational complexity increases with channel dimension, thus limiting their applicability in real-time systems \cite{b19,b20}. Several filtering-based channel tracking methods have also been developed. For example, real-time beamformer tracking is proposed in \cite{b21}, while \cite{b22} introduces a tracking approach for time- and space-varying channels based on the extended Kalman filter (EKF). In \cite{b23}, the spatial characteristics of sparse channels are captured within an SBL framework that integrates the Kalman filter (KF) and Rauch–Tung–Striebel (RTS) smoother to estimate posterior statistics during the expectation step, along with a search-based optimization process in the maximization step. More recently, deep learning (DL) techniques \cite{b24,b25} have been explored to model time-varying channel dynamics in an end-to-end manner. For example, convolutional neural networks (CNNs) \cite{b25} have been employed to infer channel parameters directly from the received signals. Despite their promising results, these DL-based approaches often rely heavily on the training data characteristics, and their ability to interpret underlying physical-layer mechanisms remains limited.

In addition, many studies have sought to enhance time-domain sparsity by extending the compressed sensing (CS) framework. For example, one approach employs a block-sparse OMP algorithm that exploits path delay correlations between adjacent symbols \cite{b26,b27}. Another adopts a two-step estimation scheme, in which the Least Squares (LS) method provides an initial coarse estimate of delay–angle parameters, followed by a CS-based refinement stage. However, under high mobility, dispersion in sparsity patterns often leads to performance degradation \cite{b17}. A different line of research introduces bi-objective optimization frameworks that jointly impose sparse and low-rank constraints, typically solved via the alternating direction method of multipliers (ADMM). While effective, such methods usually assume a fixed rank, which might potentially introduce estimation bias in rapidly changing environments \cite{b28}. Furthermore, joint spatio-temporal sparsity has been incorporated into tensor tracking algorithms based on Kalman filtering to dynamically update the low-rank structure, although computational complexity remains a limitation \cite{b15}.

While the above studies have achieved notable progress, existing matrix recovery-based estimation techniques largely rely on the assumption of a fixed channel rank, a simplification not valid in realistic mmWave environments. In reality, user mobility and environmental variations, such as cluster appearance,/disappearance, and path blockage, cause the effective channel rank to fluctuate over time. For instance, when a vehicle moves from an urban canyon into an open area, the number of dominant propagation paths may change abruptly. Conventional low-rank matrix completion (LRMC) and CS algorithms that fix the rank or sparsity level often produce inaccurate parameter estimates under such conditions. The mismatch between the assumed and actual rank propagates errors into subsequent sparse recovery stages, substantially degrading overall system performance. This issue becomes more pronounced in high-Doppler scenarios, where outdated rank assumptions lead to beam misalignment and spectral efficiency degradation. Therefore, a dynamic rank-adaptation mechanism is essential for ensuring robust channel estimation in mobile mmWave systems.

To bridge these gaps, this paper proposes a low-complexity channel estimation method for mobile mmWave systems, where the rapidly time-varying channel estimation problem is formulated as a sparse matrix recovery task, and a structured low-rank matrix sensing technique is employed to capture the channel dynamics. The main contributions of this work are summarized as follows:

\begin{itemize}

	\item \textbf{Two-stage compressed sensing framework for fast time-varying mmWave channels}: A robust rank-one matrix completion (R1MC) method is then applied to complete the observation matrix obtained from limited pilot symbols. By dynamically updating the estimated rank without increasing computational complexity, the accuracy of matrix completion is significantly enhanced. During the second stage of channel estimation, the matrix rank, under the clustered channel model, aligns with its sparsity. The known sparsity is then used as a convergence criterion in compressed sensing, and the proposed RA-BOMP algorithm estimates the channel matrix. Since RA-BOMP combines classical batch OMP with rank priors, computational complexity is substantially reduced.

	\item \textbf{Fully rank-aware channel estimation and reconstruction framework}: The framework does not assume angular stability between consecutive time frames. Instead, it utilizes rank evolution as a primary indicator of channel dynamics, enabling automatic adaptation of the estimation strategy. This supports path re-identification and gain/angle recovery, overcoming the limitations of prior angle-predictive methods.

	\item \textbf{Rank-constrained adaptive measurement matrix design (RAMMD) for improved beam alignment}: By exploiting the similarity of channel matrices across adjacent time instances and applying rank-aware control, the measurement matrix is dynamically optimized to enhance channel angle estimation, thereby improving the efficiency and accuracy of beam alignment. Using rank as a constraint and prior throughout the time-varying channel estimation process further enhances stability and accuracy while reducing computational complexity.
\end{itemize}

The remainder of this paper is organized as follows. Section \textcolor{blue}{II} introduces the system model and formulates the problem, focusing on the sparse representation of the channels in the time domain. Section \textcolor{blue}{III} proposes a two-phase compressed sensing method for channel estimation. Section \textcolor{blue}{IV} presents the design of an adaptive measurement matrix with rank awareness. Section \textcolor{blue}{V} analyzes the computational complexity of the proposed algorithm at each phase. Simulation results are provided in Section \textcolor{blue}{VI}, and Section \textcolor{blue}{VII} concludes the paper.

$\textit{Notation}$: We introduce the following notation for the remainder of this paper. Matrices are represented by the bold uppercase letter $\textbf{A}$, while column vectors are indicated by bold lowercase $\textbf{a}$, and scalars are denoted by non-bold lowercase $a$. The complex conjugate, transpose, and Hermitian transpose are represented as $(\cdot)^*$, $(\cdot)^T$, and $(\cdot)^H$, respectively. A circularly symmetric jointly Gaussian random vector with mean $\mathfrak m$ and covariance $\varsigma$ is symbolized by $\mathcal{N}(\mathfrak m, \varsigma)$. The ${{\ell }_{1}}$, ${{\ell }_{2}}$ and Frobenius norms of the vector $\textbf{a}$ are expressed as $||\textbf{a}||_1$, $||\textbf{a}||_2$ and $||\textbf{a}||_F$, respectively, $\text{rank}(\textbf{A})$ represents the rank of $\textbf{A}$.

\section{SYSTEM MODEL}
Consider a single-user frequency-selective mmWave MIMO system comprising a base station (BS) and a mobile station (MS). Assume that the BS is equipped with ${\mathcal{N}_{\text{BS}}}$ antennas and ${\mathcal{M}_{\text{BS}}}$ RF chains, while the MS is equipped with ${\mathcal{N}_{\text{MS}}}$ antennas and ${\mathcal{M}_{\text{MS}}}$ RF chains. The number of RF chains is less than the number of antennas, i.e., $\mathcal{M}_{\text{BS}} < \mathcal{N}_{\text{BS}}$ and $\mathcal{M}_{\text{MS}} < \mathcal{N}_{\text{MS}}$. The channel matrix between the BS and MS, denoted as ${{\mathbf{H}}} \in \mathbb{C}^{\mathcal{N}_{\text{MS}} \times \mathcal{N}_{\text{BS}}}$, is modeled as a frequency-selective MIMO channel. Both the BS and MS adopt a hybrid analog-digital architecture. Block transmission enables the reconfiguration of the RF circuits between consecutive time instances, thereby preventing data loss during reconfiguration and reducing inter-instance interference. The specific transmission structure is illustrated in Fig. \textcolor{blue}{1}.

\begin{figure*}[t]
	\centering %
	\includegraphics[width=16.0cm]{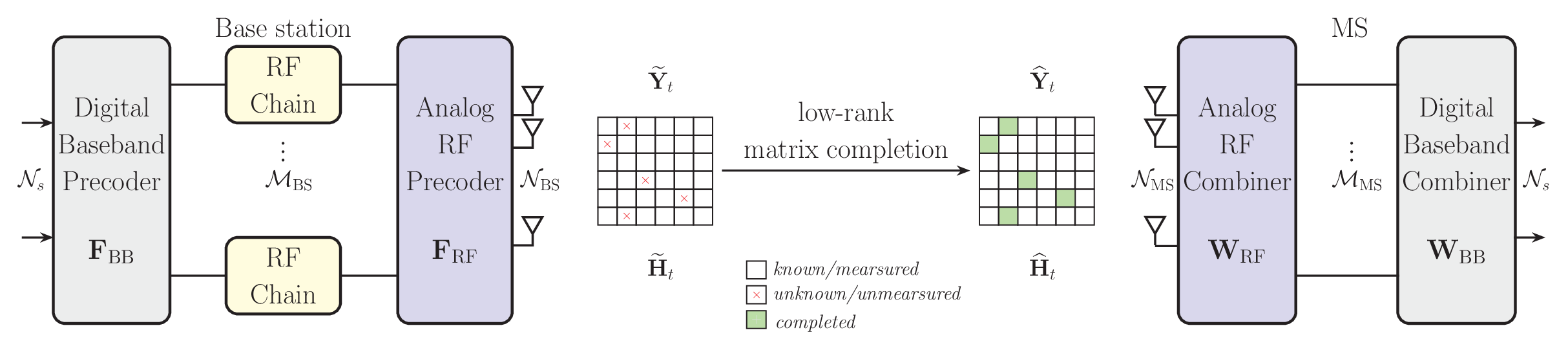}
	\caption{A schematic of mmWave MIMO signal transmission model and time-varying channel estimation and reconstruction from incomplete observations.}
\end{figure*}
\subsection{Channel Model}
Consider a clustered channel model \cite{b29,b30} for the frequency-selective mmWave MIMO channel, consisting of ${{\mathcal K}_{\ell}}$ clusters. In the beam space, the frequency-selective MIMO channel matrix ${\mathbf{H}}[d]$ with delay-$d$ can be expressed by
\begin{equation}\begin{aligned}
		\textbf{H}[d] & =\sqrt{\frac{\mathcal{N}_{\text{BS}}\mathcal{N}_{\text{MS}}}{\varrho_{\text{loss}}}}\sum_{\ell=1}^{\mathcal{L}}\sum_{\kappa_{\ell}=1}^{\mathcal{K}_{\ell}}\alpha_{\kappa_{\ell}}\varrho_{\text {rc}}\left(d\mathcal{T}_{\text{s}}-\tau_{\ell}-\tau_{\kappa_{\ell}}\right) \\
		& \times\textbf{a}_{\text{MS}}\left(\theta_{\ell}-\vartheta_{\kappa_{\ell}}\right)\textbf{a}_{\text{BS}}^{H}\left(\phi_{\ell}-\varphi_{\kappa_{\ell}}\right),
\end{aligned}\end{equation}
where $\varrho_{\text{loss}}$ is the path loss; $\mathcal{L}$ is the number of paths, and $\alpha_{\kappa_\ell}$ denotes the complex gain associated with the $\ell$th cluster. Each cluster contributes $\mathcal{K}_{\ell}$ rays/paths between the BS and MS \cite{b31,b32}. $\varrho_{\text {rc}}(\cdot)$ is a pulseshaping filter for $\mathcal T_s$-spaced signaling generated at $d\mathcal T_s-\tau_\ell-\tau_{\kappa_{\ell}}$ instance. The azimuth angles of arrival (AoAs) and departure (AoDs) of the $\ell$-th cluster are ${\theta}_{\ell} \in [0,2\pi]$ and ${\phi}_{\ell} \in [0,2\pi]$, respectively. The variables $\vartheta_{\kappa_\ell}$ and $\varphi_{\kappa_\ell}$ denote the angular offsets of the $\kappa_\ell$-th path relative to the mean AoDs and AoAs of the $\ell$-th cluster, respectively. Superscript $(\cdot)^H$ denotes the Hermitian transpose.

Assuming a uniform linear array (ULA) at both the BS and MS, the antenna array response vectors are expressed as
\begin{equation}
\begin{aligned}
& {{\textbf{a}}_{\text{MS}}}({{\theta }_{\ell}})=\frac{1}{\sqrt{{\mathcal{N}_\text{MS}}}}{[1,{{e}^{j\frac{2\pi }{\lambda }\rho\sin {{\theta }_{\ell}}}},\ldots ,{{e}^{j({{\mathcal N}_{\text{MS}}-1)\frac{2\pi }{\lambda }\rho\sin {{\theta }_{\ell}}}}]}^{{T}}}\\ 
& {{\textbf{a}}_{\text{BS}}}({{\phi }_{\ell}})=\frac{1}{\sqrt{{\mathcal{N}_\text{BS}}}}{{[1,{{e}^{j\frac{2\pi }{\lambda }\rho \sin {{\phi }_{\ell}}}},\ldots ,{{e}^{j({\mathcal{N}_\text{BS}}-1)\frac{2\pi }{\lambda }\rho \sin {{\phi }_{\ell}}}}]}^{{T}}},\\ 
\end{aligned}
\end{equation}
where $\lambda$ is the signal wavelength, and $\rho$ is the spacing between adjacent antenna elements, typically set to $\rho = \lambda/2$.

Accordingly, the frequency-selective channel model with a given elay-$d$ can be represented as \cite{b29}
\begin{equation}\textbf{H}=\sum_{d=0}^{D-1}\textbf{H}[d]e^{-j2\pi d}.\end{equation}

At time instant $t$, the channel model in (3) can be equivalently expressed as
\begin{equation}\label{Ht}
{{\textbf{H}}_t}={{\mathbf{\Theta}}_{\text{MS}}}{\bar{\textbf{H}}_t}\mathbf{\Theta}_{\text{BS}}^{H},\forall\, t=0,1,\ldots,T,
\end{equation}
where $\mathbf{\Theta}_{\text{MS}} = \{{\mathbf{a}_{\text{MS}}(\theta_\ell)}\}_{\ell=1}^{\mathcal{L}_1}$ and $\mathbf{\Theta}_{\text{BS}} = \{{\mathbf{a}_{\text{BS}}(\phi_\ell)}\}_{\ell=1}^{\mathcal{L}_2}$ are overcomplete dictionaries, and $\bar{\mathbf{H}}_t \in \mathbb{C}^{\mathcal{L}_1 \times \mathcal{L}_2}$ is a sparse matrix, i.e., the sum of $\mathcal{L}$ sparse components\footnote{It is worth noting that the methodology developed in this work can be seamlessly applied to massive MIMO-OFDM system with minimal adaptation, as the channels across OFDM subcarriers typically share the same sparse structure.}. For sparse recovery, the matrices ${\mathbf{\Theta}}_\text{BS} \in \mathbb{R}^{\mathcal{N}_{\text{BS}} \times \mathcal{L}}$ and ${\mathbf{\Theta}}_\text{MS} \in \mathbb{R}^{\mathcal{N}_{\text{MS}} \times \mathcal{L}}$ are defined and can be pre-computed at the MS. 
This clustered channel model exploits the fact that each cluster comprises multiple paths, while the low-rank structure induced by path clustering is not explicitly modeled. It can be verified that the channel matrix $\mathbf{H}_t$ has a rank equivalent to its sparsity level, assuming that grid mismatch is negligible. Therefore, the known channel rank provides a priori knowledge of sparsity that OMP can exploit. 
\subsection{Time-varying Sparse Formulation in the Time-Domain}

In this subsection, we focus on time-domain channel estimation, where the proposed framework leverages the rank–sparsity relationship to reconstruct the time-varying mmWave channel more efficiently. At the $t$-th time instance, the BS applies an RF chain $\textbf{F}_t=\textbf{F}_{\text{RF}} \textbf{F}_{\text{BB}}\in \mathbb{C}^{\mathcal{N}_\text{BS} \times \mathcal{M}_\text{BS}}$, is implemented by quantizing the angles of the analog phase shifters. The received signal at the BS, in the absence of noise, is given by

\begin{equation}\textbf{x}_t=\textbf{F}_t\textbf{s}_t,\end{equation}
where $\mathbf{s}_t \in \mathbb{C}^{\mathcal{N}_s \times 1}$ denotes the symbol vector at time instance $t$, and $N_s$ represents the number of transmittable data streams.

From (1), the unknown channel parameters within $\mathbf{H}_t$ are highly interdependent due to multipath fading, and their contributions may be partially attenuated due to non-coherent summation. Nevertheless, the signal can be represented in a structured form to facilitate sparse recovery by integrating the frame-structured channel with a specific measurement matrix. At the $t$th time instance, the MS employs an RF chain $\textbf{W}_t = \textbf{W}_\text{RF}\textbf{W}_\text{BB} \in \mathbb{C}^{\mathcal{N}_\text{MS}\times \mathcal{M}_\text{MS}}$, which is implemented using quantized phase shifts at the receiver. The resulting post-combining signal is then expressed as
\begin{equation}\textbf{Y}_t=\textbf{W}^H_t\textbf{H}_t\textbf{F}_t\textbf{S}^{{T}}_t+\textbf{z}_t,\end{equation}
where $\textbf{n}_t$ denoted additive noise vector obeying Gaussian distribution $\mathcal{N}(0,\varsigma^2\textbf{I})$. Once $M$ training symbol vectors at time instance $t$ are completed, i.e., $\textbf{s}_t=[\textbf{s}_t^T[1],\textbf{s}_t^T[2],\ldots,\textbf{s}_t^T[M]]^T$, $\textbf{y}_t=[\textbf{y}_t^T[1],\textbf{y}_t^T[2],\ldots,\textbf{y}_t^T[M]]^T$, the training and observation matrix is respectively defined as
\begin{equation}
\begin{aligned}
&\textbf{S}_{t}=[\textbf{s}_t[:,1],\textbf{s}_t[:,2],\ldots,\textbf{s}_t[:,\mathcal{N}_\text{BS}]],\\
&\textbf{Y}_{t}=[\textbf{y}_t[:,1],\textbf{y}_t[:,2],\ldots,\textbf{y}_t[:,\mathcal{N}_\text{BS}]].
\end{aligned}
\end{equation}

In this paper, we implement the precoder and combiner using high-resolution phase shifters (PSs) to realize analog beamforming. For notational convenience, we define $(\textbf{F}_t^T\textbf{S}_t)\otimes\textbf{W}_t^H$ by ${\bm{\mathcal X}}_t$, so that (6) can be equivalently rewritten as
\begin{equation}\begin{aligned}
\textbf{Y}_{t} & =\textbf{W}_t^H\textbf{H}_t\textbf{F}_t\textbf{S}_t+\textbf{Z}_t \\
& =\left(\textbf{F}_t\textbf{S}_t\right)^T\otimes\textbf{W}_t^H\textbf{H}_t+\textbf{Z}_t \\
& \triangleq{\bm{\mathcal X}}_t\textbf{H}_t+\textbf{Z}_t,
\end{aligned}\end{equation}
where $\otimes$ denotes the Kronecker product, and ${\bm{\mathcal X}}_t$ serves as the measurement matrix (constructed dictionary). The matrices $\mathbf{Y}_t$ and $\bf{H}_t$ are reshaped to match their original dimensions. Assuming a normalized symbol set satisfying $||\mathbf{S}_t|| = \mathbf{I}$, we have ${\bm{\mathcal X}}_t = \mathbf{F}_t^T \otimes \mathbf{W}_t^H$.

Since the observation matrix may be incomplete or corrupted, channel estimation from such observations can be formulated as a matrix completion problem, i.e., a sparse signal recovery problem.

\subsection{Problem Formulation}

\begin{figure}[t]
	\centering %
	\includegraphics[width=8cm]{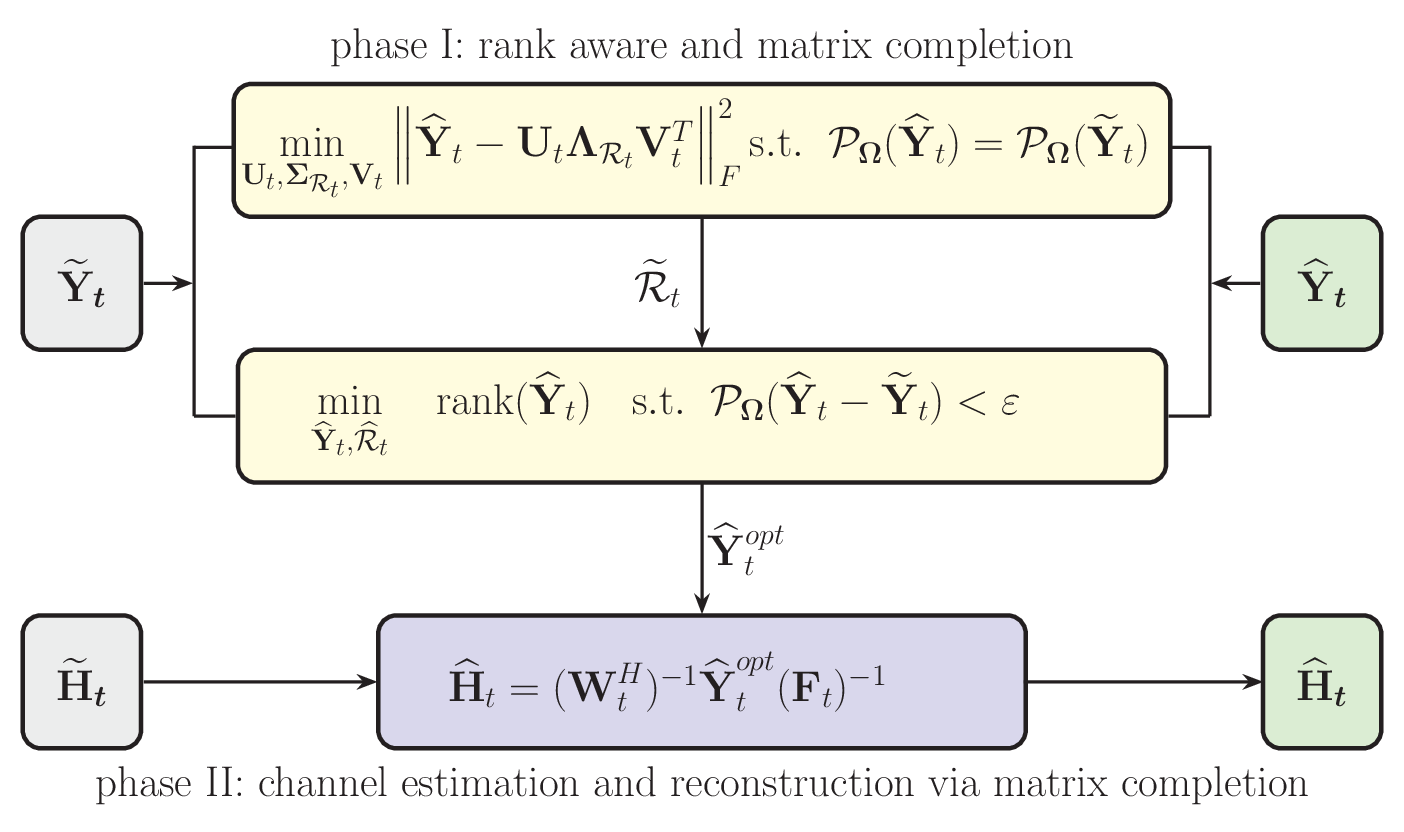}
	\caption{Block diagram illustrating the proposed signal processing framework.}
\end{figure}

In mmWave frequency bands, the number of significant propagation paths $\mathcal{L}$ is typically small. Hence, the observation model in (4) can be formulated as a sampling process of a low-rank matrix. With imperfect CSI, the incomplete observation is mathematically expressed as \cite{b2}:
\begin{equation}(\widetilde{\textbf{Y}}_t)_{ij}=(\textbf{W}_t^H\textbf{H}_t\textbf{F}_t)_{ij}\quad(i,j)\in{\mathbf\Omega},\end{equation}
where $(\widetilde{\mathbf{Y}}_t)_{ij}$ denotes the $(i,j)$-th entry of $\widetilde{\mathbf{Y}}_t$, and $\mathbf\Omega$ denotes the index set of observed entries.

Exploiting this intrinsic low-rank property, the observation matrix completion problem can be posed as a low-rank matrix completion (LRMC) problem:
\begin{equation}\min_{\widehat{\textbf{Y}}_t}\quad\operatorname{rank}(\widehat{\textbf{Y}}_t)\quad\operatorname{s.t.}\|\mathcal{P}_{\boldsymbol{\Omega}}(\widehat{\textbf{Y}}_t-\widetilde{\textbf{Y}}_t)\|_F^2<\varepsilon,\end{equation}
where $\vert\vert\cdot\vert\vert_F$ denotes the Frobenius norm, and $\widetilde{\mathbf{Y}}_t$ is the incomplete observation matrix. Here, $\mathcal{P}_{\boldsymbol{\Omega}}$ denotes the sampling operator projecting onto the observed entries, and $\boldsymbol{\Omega} \in \mathbb{R}^{\mathcal{L}_1 \times \mathcal{L}_2}$ is a binary matrix indicating the sampling domain.

Specifically, the low-rank matrix $\widehat{\mathbf{Y}}_t^{opt}$ can be efficiently reconstructed via nuclear-norm minimization:

\begin{equation}\widehat{\textbf{Y}}_t^{opt}\triangleq\arg\min_{\widehat{\textbf{Y}}_t}\quad\|\mathcal{P}_{\boldsymbol{\Omega}}(\widehat{\textbf{Y}}_t)-\mathcal{P}_{\boldsymbol{\Omega}}(\widetilde{\textbf{Y}}_t)\|_F^2+\sigma\|\widehat{\textbf{Y}}_t\|_*,\end{equation}
where $\sigma$ is a positive regularization parameter, which can be determined via singular value decomposition (SVD); $|\widehat{\textbf{Y}}_t\|_*=\text {Tr}\left (\sqrt{\widehat{\textbf{Y}}_t^T\widehat{\textbf{Y}}_t}\right )$, which is equivalent to the tightest convex relaxation of the rank minimization.

Subsequently, the refined channel matrix $\widehat{\mathbf{H}}_t$ is estimated as
\begin{equation}\widehat{\textbf{H}}_t=(\textbf{W}_t^H)^{-1}\widehat{\textbf{Y}}_t^{opt}(\textbf{F}_t)^{-1}.\end{equation}

It is noted that, as in (11), before the observation matrix $\widehat{\mathbf{Y}}_t$ is stably completed, only a coarse channel estimate $\widetilde{\textbf{H}}_t=(\textbf{W}_t^H)^{-1}\widetilde{\textbf{Y}}_t(\textbf{F}_t)^{-1}$ can be obtained. Subsequently, the channel estimation under incomplete observations will be addressed in two consecutive stages.

\section{Two-Phase Compressed Sensing Scheme}

It is commonly assumed that, within a single time instance, the angles of arrival (AoAs) and angles of departure (AoDs) of time-varying mmWave channels remain constant. However, this assumption may lead to significant inaccuracies in channel estimation. Therefore, a dynamic adaptation mechanism is required to accurately track the evolution of propagation paths, including cases where paths appear, disappear, or undergo angular shifts, thereby improving both the stability and accuracy of the channel estimates. In this Section, using the model illustrated in Fig. \textcolor{blue}{2}, we estimate and reconstruct the time-varying channels at each time instance in two consecutive phases.
 
\subsection{Robust Rank-one Matrix Pursuit and Dynamic Rank Estimation}
Additionally, many existing approaches rely on a pre-specified rank, which can significantly affect algorithm performance. To address this issue, we propose a robust rank-one matrix completion (R1MC) method with integrated dynamic rank estimation. In R1MC, the data matrix is represented as a weighted sum of rank-one matrices, and the rank is estimated based on these weights. This approach provides high-resolution information, enabling the reconstruction of a complete and accurate signal $\widehat{\mathbf{Y}}_t$ from a limited number of pilot symbols. 
1) $\textit{Dynamic Rank Estimation}$: To estimate the rank of $\widehat{\mathbf{Y}}_t$, we follow the approach proposed in \cite{b10}, expressed as
\begin{equation}
\min_{\textbf{U}_t,\mathbf{\Sigma}_{\widetilde{\mathcal{R}}_t},\textbf{V}_t}\|\widehat{\textbf{Y}}_t-\underbrace{\textbf{U}_t\mathbf{\Sigma}_{\widetilde{\mathcal{R}}_t}\textbf{V}_t^T}_{\widetilde{\textbf{Y}}_t}\|_F^2\quad\mathrm{s.t.}\mathcal{P}_{\mathbf{\Omega}}(\widehat{\textbf{Y}}_t)=\mathcal{P}_{\mathbf{\Omega}}(\widetilde{\textbf{Y}}_t),\end{equation}
where the factor matrices $\mathbf{U}_t \in \mathbb{C}^{\mathcal{L}_1 \times \widetilde{\mathcal{R}}_t}$ and $\mathbf{V}_t \in \mathbb{C}^{\widetilde{\mathcal{R}}_t \times \mathcal{L}_2}$ have orthogonal columns. $\mathbf{\Sigma}_t \in \mathbb{C}^{\widetilde{\mathcal{R}}_t \times \widetilde{\mathcal{R}}_t}$ is a diagonal matrix, which computes the best rank-$\widetilde{\mathcal{R}}_t$ approximation of the trimmed matrix via sparse SVD. The rank $\widetilde{\mathcal{R}}_t$ is determined as the index corresponding to the minimum ratio between two consecutive singular values. $\mathcal{L}_1$ and $\mathcal{L}_2$ denote the dimensions of the overcomplete grid in the angle domain.

Generally, $\widehat{\mathbf{Y}}_t$ can be represented as the weighted sum of $\widetilde{\mathcal{R}}_t$ rank-one matrices:
\begin{equation}\widetilde{\textbf{Y}}_t=\textbf{U}_t\mathrm{diag}(\boldsymbol{\sigma})\textbf{V}_t^T=\sum_{i=1}^{\widehat{\mathcal{R}}_t}\sigma_i\cdot\textbf{u}_i\textbf{v}_i,\end{equation}
where the singular values $\sigma_i$ are sorted in descending order, i.e., $\sigma_1 \ge \sigma_2 \ge \cdots \ge \sigma_{\widehat{\mathcal{R}}_t}$, and $\mathbf{u}_i$ and $\mathbf{v}_i$ denote the $i$th left and right singular vectors, respectively. Note that SVD provides a rank-one approximation through its singular vectors and values. The weight vector $\boldsymbol{\sigma}$ obtained from (14) is further refined under rank-awareness to enforce low-rank constraints.

Following (\textcolor{blue}{13}), the effective rank of $\widetilde{\mathbf{Y}}_t$ is estimated as $\widehat{\mathcal{R}}_t$. Let $\{{\sigma_i}\}_{i=1}^{\widehat{\mathcal{R}}_t}$ denote the singular values in descending order. The effective rank is determined by
\begin{equation}\label{rank}
\begin{aligned}\widehat{\mathcal{R}}_t=\max\left\{k\in\mathbb{Z}^+:\sum_{i=1}^k\sigma_i\geq\xi_k\sum_{i=1}^n\sigma_i\right\},\end{aligned}\end{equation}
where $\xi_k \in (0,1)$ controls the energy retention ratio. The estimated rank $\widehat{\mathcal{R}}_t$ determines both the sparsity level in orthogonal matching pursuit (OMP) and the dimension of the measurement matrix ${\bm{\mathcal X}}_t \in \mathbb{C}^{(\mathcal{N}_{\mathrm{MS}}\mathcal{N}_{\mathrm{BS}}) \times \widehat{\mathcal{R}}_t}$.
To address abrupt rank changes caused by user mobility, an online rank predictor is incorporated into R1MC. The rank sequence ${\widehat{\mathcal{R}}_t}$ is modeled as a Markov process and approximated with a discrete-time autoregressive (AR) process of order $j$:
\begin{equation}\label{AR}\widehat{\mathcal{R}}_t=\sum_{j=1}^ja_j\widehat{\mathcal{R}}_{t-j}+s_f\mathcal Z_t,\end{equation}
where $a_j$ denotes the $j$th weighting coefficient, $s_f$ is a scaling factor, and $\mathcal Z_t \sim \mathcal{N}(0,1)$ are independent and identically distributed (i.i.d.). The rank estimates ${\widehat{\mathcal{R}}_t}$ are obtained via the following procedure: initially, it is assumed that the rank estimate can be set as $\widehat{\mathcal{R}}_{t-1}$, which is typically used to initialize the matrix reconstruction process.

2) $\textit{Matrix Completion}$: Once the rank estimation is completed, we proceed with matrix completion of the incomplete matrix $\mathcal{P}_\mathbf{\Omega}$ indexed by the set $\mathbf{\Omega}$. Recalling (\textcolor{blue}{14}), this process can be regarded as a specific low-rank matrix decomposition for completion, i.e., rank-one matrix completion (R1MC). Let $||\textbf{u}_i||_2=1$ and $||\textbf{v}_i||_2=1$ for $i=1,\ldots,{\widehat{\mathcal{R}}_t}$, such that $\textbf{u}_i$ and $\textbf{v}_i$ form orthonormal bases in $\mathbb{R}^{n}$, respectively. By applying a generalized least absolute shrinkage and selection operator (Lasso), or equivalently, an $\ell_{1}$-norm regularization on $\boldsymbol{\sigma}$, the Lasso-type matrix recovery problem can be formulated as

\begin{equation}\begin{aligned}\min_{\boldsymbol{\sigma},\{\textbf{u}_{i},\textbf{v}_{i}\}_{i=1}^{\widehat{\mathcal{R}}_{t}}}&\frac{1}{2}\|\widehat{\textbf{Y}}_{t}-\sum_{i=1}^{\widehat{\mathcal{R}}_{t}}\sigma_{i}\textbf{u}_{i}\textbf{v}_{i}^{T}\|_{F}^{2}+\mu\|\boldsymbol{\sigma}\|_{1}\\\mathrm{s.t.}&\mathcal{P}_{\boldsymbol{\Omega}}(\widehat{\textbf{Y}}_{t})=\mathcal{P}_{\boldsymbol{\Omega}}(\widetilde{\textbf{Y}}_{t}),\end{aligned}\end{equation}
where $\mu$ is an augmented Lagrangian parameter and $\boldsymbol{\sigma}= (\sigma_1,\ldots,\sigma_{\widehat{\mathcal{R}}_t})$.

To efficiently solve this multi-dimensional optimization problem, the alternating direction method of multipliers (ADMM) \cite{b33} is adopted. For simplicity, we denote $\mathcal{B}_i=\{\sigma_i,\textbf{v}_i,\textbf{u}_i\}, i=1,\ldots,\widehat{\mathcal{R}}_t$ as the $i$-th block coordinate descent variable. With the augmented Lagrangian function, (17) is converted into an unconstrained minimization problem given by
\begin{equation}\begin{aligned}\mathscr{L}(\textbf{Y},\boldsymbol{\sigma},\textbf{v}_{i},\textbf{u}_{i};\mathbf{\Lambda})&=\frac{1}{2}||\widehat{\textbf{Y}}_{t}-\sum_{i=1}^{\widehat{\mathcal{R}}_{t}}\sigma_{i}\cdot\textbf{v}_{i}\textbf{u}_{i}^{T}||_{F}^{2}+\\&\operatorname{tr}(\mathbf{\Lambda}^{H}(\widehat{\textbf{Y}}_{t}-\sum_{i=1}^{\widehat{\mathcal{R}}_{t}}\sigma_{i}\cdot\textbf{v}_{i}\textbf{u}_{i}^{T}))+\mu\parallel\boldsymbol{\sigma}\parallel_{1},\end{aligned}\end{equation}
where $\mathbf{\Lambda}\in\mathbb{R}^{\mathcal{L}_1\times\mathcal{L}_2}$ is the Lagrange multiplier and $\mu>0$ is the penalty coefficient.
In standard ADMM iterations, $\textbf{Y}$ and $\mathbf{\Lambda}$ are fixed first, and $\{{\textbf{v}_i,\textbf{u}_i}\}_{i=1}^{\widehat{\mathcal{R}}_t}$ and $\boldsymbol{\sigma}$ are updated sequentially. Consequently, (18) can be simplified as
\begin{equation}\mathscr{L}(:,\mathcal{B}_i;:)=\frac{1}{2}\left|\left|\textbf{Y}_i-\sigma_i\cdot\textbf{v}_i\textbf{u}_i^T\right|\right|_F^2+\mu|\sigma_i|,\end{equation}
where $\textbf{Y}_i=\widehat{\textbf{Y}}_t-\sum_{i=1}^{\widehat{\mathcal{R}}_t}\sigma_i\cdot\textbf{v}_i\textbf{u}_i^T$ is the residual of the approximation.

To strike a balance between the accuracy of reconstructed channel and the computational complexity of the CSI acquisition, we introduce a block coordinate descent (BCD) method, where ${\textbf{v}_i,\textbf{u}_i,\sigma_i}$ are treated as $\widehat{\mathcal{R}}_t$ independent blocks. Thus, each block $\mathcal{B}_i^{(n+1)}$, defined by a series of iterative shrinkage-thresholding, is updated by
\begin{equation}\mathcal{B}_i^{(n+1)}=\left\{shrinkage_\mu(\nu),\frac{\textbf{Y}_i^{T^{(n)}}\textbf{u}_i^{(n+1)}}{\sigma_i},\frac{\textbf{Y}_i^{(n)}\textbf{v}_i^{(n)}}{\sigma_i}\right\},\end{equation}
where $\nu = \langle\textbf{Y}_i^{(n)}, \textbf{v}_i^{(n+1)}(\textbf{u}_i^T)^{(n+1)}\rangle$, and the soft-shrinkage operator is defined as ${sign}(\nu)\max(\vert \nu\vert-\mu,0)$.

With the help of ADMM iterations, the proposed framework can handle matrix completion in a unified manner, i.e., solving
\begin{equation}\widehat{\textbf{Y}}_t^{opt}=\arg\min||\widehat{\textbf{Y}}_t||_*\quad\mathrm{s.t.}||\mathcal{P}_{\boldsymbol{\Omega}}(\textbf{Y}_t)-\mathcal{P}_{\boldsymbol{\Omega}}(\widehat{\textbf{Y}}_t)||_F\leq\varepsilon.\end{equation}

Finally, when $\mathcal{B}_i$ is obtained from (\textcolor{blue}{20}), the estimated rank $\widehat{\mathcal{R}}_t$ constrains the subsequent channel estimation and the design of the measurement matrix, as summarized in Algorithm \textcolor{blue}{1}. Upon completing the observation matrix estimation, the second phase proceeds with the reconstruction of the channel matrix.

\begin{algorithm}
	\caption{Rank-One Matrix Completion (R1MC)}
	\KwIn{$\widetilde{\textbf{Y}}_{t}$, $\mathbf{\Omega}$, $\mathcal R$}
	$\textbf{Initialization}$: {$\mathcal{P}_\mathbf{\Omega}(\widetilde{\textbf{Y}}_{t}) = \mathcal{P}_\mathbf{\Omega}(\widehat{\textbf{Y}}_{t})$, $\mathcal{P}_{\mathbf{\Omega}^c}(\widetilde{\textbf{Y}}_{t}) = 0$, $\textbf{Z} = \text{zeros}(I_1, I_2)$}\\
	\For{$kn= 0, \ldots, N-1$}{\qquad\qquad\qquad\qquad\qquad\qquad\qquad\quad$\triangleright \,${iterations $N$}\\
		\textbf{Compute} approximation $\widehat{\mathcal{R}}_t$ of $\widetilde{\textbf{Y}}_{t}$ by \eqref{rank} and \eqref{AR}\\
		\quad $[\textbf{U}_t^{(n)} \mathbf{\Sigma}_t^{(n)} \textbf{V}_t^{(n)}] \gets \text{SVD}(\widetilde{\textbf{Y}}_{t})$\\
		\quad $\textbf{U}_t^{(n)} = \{ \textbf{u}_i \}_{i=1}^{\widehat{\mathcal{R}}_t} \in \mathbb{C}^{\mathcal{L}_1\times\widehat{\mathcal{R}}_t}$ \\
		\quad $\mathbf{\Sigma}_t^{(n)} \in \mathbb{C}^{\widehat{\mathcal{R}}_t\times\widehat{\mathcal{R}}_t}\gets \sigma_1,..., \sigma_{\widehat{\mathcal R}_t}$ \\
		\quad $\textbf{V}_t^{(n)} = \{\textbf{v}_i \}_{i=1}^{\widehat{\mathcal{R}}_t} \in \mathbb{C}^{\widehat{\mathcal{R}}_t\times\mathcal{L}_2}$\\
		\textbf{Set} $\textbf{Z} = \textbf{U}_t^{(n)} \mathbf{\Sigma}_t^{(n)} ({\textbf{V}_t^{T}})^{(n)}$\\
		\textbf{Update} missing entries: $\mathcal{P}_{\mathbf{\Omega}^c}(\widetilde{\textbf{Y}}_{t}) = \mathcal{P}_{\mathbf{\Omega}^c}(\widehat{\textbf{Y}}_{t})$\\
		$\textbf{If}$ {$\frac{\| \mathcal{P}_\mathbf{\Omega}(\widetilde{\textbf{Y}}_{t} - \widehat{\textbf{Y}}_{t}) \|_F}{\| \mathcal{P}_\mathbf{\Omega}(\widetilde{\textbf{Y}}_{t}) \|_F} < \epsilon$ \textbf{or} $\frac{\| {\widetilde{\textbf{Y}}_{t}}^{n+1} - {\widetilde{\textbf{Y}}_{t}}^{(n)} \|_F}{\| {\widetilde{\textbf{Y}}_{t}}^{(n+1)} \|_F} < \epsilon$}\\
		\qquad\qquad\qquad\qquad\qquad\qquad\qquad\qquad$\triangleright \,${tolerance $\epsilon$}\\
		\quad	\textbf{break}
	}
	\KwOut{$\widehat{\textbf{Y}}_{t}$}
\end{algorithm}

\subsection{Phase II: Channel Estimation and Reconstruction via Completed Observations}

In Phase II, the objective is to reconstruct the underlying channel parameters $\textbf{H}_t$ from the completed observation matrix $\widehat{\textbf{Y}}_t$ obtained in the first phase.
Let the set of parameters to be estimated be denoted as $\mathcal{C} =\{ {\boldsymbol{\theta}, \textbf{g}, \boldsymbol{\tau} }\}$, representing the angles of arrival (AoAs), path gains, and path delays, respectively. The reconstruction process leverages both the low-rank and sparse characteristics of mmWave MIMO channels in the virtual angular–delay domain.
Due to the sparse scattering nature of mmWave propagation, $\textbf{H}_t$ admits a sparse representation over a structured dictionary constructed from the transmit and receive array responses. Under ideal dictionary alignment, the rank of $\textbf{H}_t$ directly corresponds to the number of effective propagation paths $\mathcal{L}$, such that $\mathrm{rank}(\textbf{H}_t) \approx \mathcal{L}$. The rank estimate $\widehat{\mathcal{R}}_t$ obtained during the completion stage thus provides an upper bound on the sparsity level of the virtual channel matrix $\bar{\textbf{H}}_t$, thereby eliminating the need for empirical sparsity tuning and enhancing estimation stability.

The virtual beamspace measurement process is modeled using the dictionary $\bm{\mathcal D}_t = \mathbf{\Theta}_{{\text{MS}}} \otimes \mathbf{\Theta}_{\text{BS}}^H$.
Accordingly, the reconstruction of $\bar{\textbf{H}}_t$ can be formulated as a a rank-constrained least-squares problem:
\begin{equation}\bar{\textbf{H}}_t^{opt}= \arg\min_{\bar{\textbf{H}}} \left\| \widehat{\textbf{H}}_t - \bm{\mathcal D}_t \bar{\textbf{H}}_t \right\|_F^2, \quad \text{s.t. } \text {rank}(\bar{\textbf{H}}_t) \leq \widehat{\mathcal{R}}_t.\end{equation}
The sparsity pattern of $\bar{\textbf{H}}_t^{opt}$ identifies the dominant angular–delay components corresponding to nonzero propagation paths.

Once $\bar{\textbf{H}}_t^{opt}$ is obtained, the complete channel matrix is reconstructed as
\begin{equation}\widehat{\textbf{H}}_t = (\mathbf{\Theta}_{\text{MS}}^H)^{-1} \bar{\textbf{H}}_t^{opt} \widehat{\textbf{Y}}_t^{opt} \mathbf{\Theta}_{\text{BS}}.\end{equation}

To efficiently address the sparse recovery problem, the Rank-Aware Batch Orthogonal Matching Pursuit (RA-BOMP) algorithm is employed.
Unlike conventional OMP, RA-BOMP incorporates the estimated rank $\widehat{\mathcal{R}}_t$ as an explicit sparsity constraint.
This enables the algorithm to terminate exactly after $\widehat{\mathcal{R}}_t$ iterations, eliminating the need for residual norm monitoring or heuristic thresholds and reducing redundant computations.

At each iteration, RA-BOMP needs to identify the column (or element) in the dictionary that can maximize the projection onto the current residual signal.  The residual $\bm{\mathfrak r}^{(n)}$ is defined as the difference between the observed signal $\widehat{\textbf{H}}_t^{(n)}$ and the reconstructed component based on the previously selected elements ${\bm{\mathcal X}}_t\bar{\textbf{H}}_t^{(n)}$, i.e., $\bm{\mathfrak r}^{(n)}=\widehat{\textbf{H}}_t^{(n)}-{\bm{\mathcal X}}_t\bar{\textbf{H}}_t^{(n)}$.

Following the sparse approximation algorithms proposed in \cite{b34}, sparse signals can be effectively recovered from a set of random linear measurements. To further enhance estimation accuracy and convergence speed, a popular sparse approximation technique, e.g., generalized OMP, can be employed to identify the best-fit projections of multidimensional observations onto the span of an overcomplete dictionary. The set comprising the indices of all non-zero elements in ${\bf H}_t$ is termed the support of ${\bf H}_t$, denoted by $\text{supp}({\bf H}_t)\triangleq\{i|[{\bf H}_t]_i\neq 0\}$,  which is determined one by one and in each iteration.

Accordingly, the support set is iteratively updated as \cite{b37}
\begin{equation}\mathcal J^{(n+1)} = \mathcal J^{(n)} \cup \mathcal{D}^{(n)} ,\end{equation}
where $\mathcal{D}^{(n)}=\{d_1,...d_N\}$ denotes the $N$ indices as of the newly selected elements. 

The coefficient update is performed by a least-squares projection onto the span of the selected elements:
\begin{equation}\bar{\textbf{H}}_{t,\mathcal J^{(n+1)}} = \arg\min_{\bar{\textbf{H}}_t^{opt}} \left\| \widehat{\textbf{H}}_t - \bm{\mathcal D}_{t,\mathcal J^{(n+1)}}\bar{\textbf{H}}_t\right\|_F^2,\end{equation} 
and the residual ${\bm {\mathfrak r}}^{(n+1)}$ is updated by subtracting $\bm{\mathcal D}_{t,\mathcal J^{(n+1)}}\bar{\textbf{H}}_{t,\mathcal J^{(n+1)}}$ from the measurements $\widehat{\textbf{H}}_t$. For a detailed description of the generalized OMP procedure, readers may refer to \cite{b37}.

The rank-aware termination criterion improves both computational efficiency and estimation robustness.
Since the effective rank of mmWave channels tends to remain stable across adjacent time instances, the computational load can be further reduced through a gain-only update mode.
Specifically, when $\widehat{\mathcal{R}}_t = \widehat{\mathcal{R}}_{t-1}$ and the angular supports exhibit negligible variation, only the gain vector $\textbf{g}_t$ is updated using the latest observations.

This assumption is consistent with the physical characteristics of mobile propagation, where angular variations evolve much more slowly than amplitude fluctuations.
Once the support sets have been identified, the estimation of physical parameters in $\mathcal{C}$ follows directly.
The nonzero indices in $\bar{\textbf{H}}_t^{opt}$ correspond to quantized AoA/AoD directions and delay bins, while the associated complex amplitudes represent the path gains.
Hence, the sparse recovery process naturally completes the parametric channel estimation task.

\section{Rank-Aware Measurement Matrix Design}

In this Section, we propose a novel rank-aware adaptive measurement matrix design method inspired by the narrow angular spread within each cluster. This algorithm adaptively designs the measurement matrix to enhance the resolution of AoAs and AoDs. Owing to the temporal correlation of the channels, its angular support set also exhibits temporal consistency, implying that the angular variations are slow over a few consecutive frames. Consequently, adjacent time instances tend to share a significant fraction of angular components and exhibit similar AoA/AoD pairs and channel ranks across successive frames. Therefore, the measurement matrix for the next time instance is designed based on the angular information from the previous time instances, effectively constraining the angular search space.

We focus the measurement matrix on specific angular regions through beamforming. Based on the angle information from the previous $t-1$ time instances, the beam direction is adjusted to point towards potential signal arrival directions, focusing on certain angular regions and reducing measurements for irrelevant angles. 

In the angle quantization of the mmWave channel, the estimated time-varying channel is projected onto a predefined angular grid. The corresponding energy distribution over the quantized angles can be expressed as
\begin{equation}
	{{\boldsymbol{\eta}}_{[\ell]}}=||\hat{\textbf{H}}_t^{H}{\bm{\mathcal D}}_{t,{[:,\ell]}}|{{|}_{2}}, \ell=1,2,\ldots ,{\mathcal{L}_1},
\end{equation}
where ${\bm{\mathcal D}}_{t,{[:,\ell]}}$ denotes the $\ell$th column of $\bm{\mathcal D}_t$. Thus, the energy corresponding to each AoA can be obtained as $\sum_{\ell=1}^{\mathcal{L}_1} {\boldsymbol{\eta}}_{[\ell]}$, representing the angular-domain power profile of the estimated channel. 

Following the results of Phase I, the number of dominant channel paths $\mathcal R_t$ (i.e., the channel rank) is determined, and the virtual angles are sorted in descending order of energy.
The total $\widehat{\mathcal{R}}_{t}$ angles are selected as cluster centroids, where the expected AoAs $\theta_i$ ($i = 1, 2, \ldots, \widehat{\mathcal{R}}_t$) fall within the angular regions of their respective clusters.

The estimated AoA range is defined as
\begin{equation}
{{{\tilde{\theta}}}_{\ell}}=( \hat{\theta}^{\max }_{i}- \hat{\theta }^{\min }_{i})\frac{\ell}{\mathcal{L}_1}+ \hat{\theta}^{\min }_{i},\, i = 1, 2, \ldots, \widehat{\mathcal{R}}_t,
\end{equation}
where $\tilde{\theta}_{\ell}$ is uniformly distributed within $[\hat{\theta}^{\min }_{i}, \hat{\theta}^{\max }_{i}]$. Generally, based on the signals received at a multi-antenna BS, efficient AOA estimation can be performed by the classic multiple signal classification (MUSIC) algorithm.

The receive steering matrix for the $i$th cluster is recalculated as
 \begin{equation} \mathbf{\Theta}_{\text{MS},[i] }= [\textbf{a}_\text{MS}(\tilde{\theta}_1), \textbf{a}_\text{MS}(\tilde{\theta}_2), \dots, \textbf{a}_\text{MS}(\tilde{\theta}_{\mathcal L_1})], 
 \end{equation}
and the corresponding dictionary component for the next time instance is updated as
 \begin{equation}
\bm{\mathcal D}_{t+1,{[{:,i}]}}=\mathbf{\Theta}^{*}_\text{BS} \otimes \mathbf{\Theta}_{\text{MS},[i]}.
\end{equation}
By adaptively steering the beamforming directions, the new measurement matrix is optimized to cover the target angular range. Specifically, the measurement matrix ${\bm{\mathcal X}}_{t+1}$ is constructed via the antenna array response vectors corresponding to the updated beam directions, thereby achieving a more compact sparse representation.

It is composed of submatrices ${\bm{\mathcal X}}_{t+1,[i]}=\bm{\mathcal D}^*_{t+1,{[{:,i}]}},\quad{{\theta}_{i}}\in [\hat{\theta}_{i}^{\min },\hat{\theta }_{i}^{\max }]$ for the $i$th cluster, each corresponding to one cluster. The post-design objective follows a dual strategy: the first component focuses on the most promising directions identified from the previous time instance, while the second introduces randomized directions to enhance robustness and mitigate the risk of missing potential signal paths.

Thereafter,  the new approximation of the estimation and the new residual can be calculated by
\begin{equation}
\begin{aligned}
\widehat{\textbf{H}}_t^{(n+1)}&=\bm{\mathcal X}_{t+1}\bar{\bf Y}_t\\
	\bm{\mathfrak r}^{(n+1)}&=\widehat{\textbf{H}}_t^{(n+1)}-{\bm{\mathcal X}}_{t+1}\bar{\textbf{H}}_t^{(n+1)}, 
	\end{aligned}
\end{equation}
where $\bar{\bf Y}_t=\arg\min_{\bar{\bf Y}_t}\vert\vert \bar{\bf Y}_t^{opt}-\bm{\mathcal X}_{t+1}\bar{\bf Y}_t\vert\vert$. Finally, the identified dominant directions are primarily used for beam reception, and randomized phase vectors are applied to out-of-range regions to enhance robustness and improve the likelihood of detecting previously unidentified angular components \cite{b35}.

\section{Analysis of Convergence and Complexity}

To ensure the robustness and reliability of the proposed two-phase channel estimation framework, we provide a rigorous theoretical analysis comprising both convergence guarantees for the R1MC algorithm and reconstruction conditions for the low-rank matrix completion (LRMC) problem in dynamic estimation.

\subsection{Convergence of R1MC Algorithm}

The R1MC algorithm in Phase I employs the ADMM framework to solve the optimization problem in (18). We establish that the ADMM iterations converge to a stationary point under standard regularity conditions, thereby guaranteeing reliable matrix completion.

Consider the augmented Lagrangian function defined in (\textcolor{blue}{19}). Let the observation matrix \(\tilde{\textbf{Y}}_t \in \mathbb{C}^{\mathcal{N}_\text{MS}
  \times \mathcal{N}_\text{BS} }\) have rank \(\mathcal{R}_t\), and assume the sampling operator \(\mathcal{P}_{\mathbf\Omega}\)  satisfies the restricted isometry property (RIP) with constant \(\delta_{\mathcal{R}_t} < 1\). If the penalty parameter \(\mu > 0\) is sufficiently large, the ADMM iterations for (\textcolor{blue}{19}) converge to a stationary point of the optimization problem (18).

Generally, the convergence result follows from the convexity of the \(\ell_1\)-norm regularization term \(\mu \|\boldsymbol{\sigma}\|_1\) and the quadratic structure of the Frobenius norm term. Define the primal variable \({\nu} = \sum_{i=1}^{\hat{\mathcal{R}}_t} \sigma_i \cdot \textbf{v}_i \textbf{u}_i^T\). In (18), the ADMM updates alternately minimize the augmented Lagrangian \(\mathscr{L}\) with respect to $\nu$, $\boldsymbol{\sigma}$, and $\mathbf{\Lambda}$. The subproblem for each pair \(\textbf{v}_q, \textbf{u}_q\) (\textcolor{blue}{21}) is strongly convex, and the soft-shrinkage operator ensures a unique closed-form solution for \(\sigma_q\). Following the convergence results of ADMM in \cite{b36}, the sequence $\{\nu^{(n)}, \boldsymbol{\sigma}^{(n)}, \mathbf{\Lambda}^{(n)}\}$ remains bounded, and the augmented Lagrangian \(\mathscr{L}\) decreases monotonically over iterations. Under the RIP assumption, the iterates converge to a stationary point that satisfies the Karush-Kuhn-Tucker (KKT) conditions.

The convergence rate is primarily influenced by the penalty parameter \(\mu\) and the condition number of the observation matrix \(\tilde{\textbf{Y}}_t\). In practice, we set \(\mu = 10^{-3} \sqrt{\mathcal{N}_\text{MS}\mathcal{N}_\text{BS} }\) to achieve a balance between convergence speed and numerical stability.
\subsection{Reconstruction Guarantees for LRMC}

The LRMC problem in (11) seeks to reconstruct a low-rank matrix \(\widetilde{\textbf{Y}}_t^*\) from incomplete observations \(\mathcal{P}_{\mathbf\Omega}(\widetilde{\textbf{Y}}_t)\). We establish the sufficient conditions for exact econstruction by leveraging the nuclear-norm minimization and the restricted isometry property. Suppose \(\widetilde{\textbf{Y}}_t \in \mathbb{C}^{\mathcal{N}_\text{MS}
  \times \mathcal{N}_\text{BS} }\) is a rank-\(\widehat{\mathcal{R}}_t\) matrix with \(\mathcal{N}_\text{MS}
 , \mathcal{N}_\text{BS}
  \gg \mathcal{R}_t\), and the sampling operator \(\mathcal{P}_{\mathbf\Omega}\) is uniformly random with \(|\mathbf\Omega|\) observed entries. Then, the solution to (\textcolor{blue}{11})
recovers \(\tilde{\textbf{Y}}_t\) exactly with probability at least \(1 - \delta\), provided:
\[
|\mathbf\Omega| \geq \zeta \mathcal{R}_t (\mathcal{N}_\text{MS}
  + \mathcal{N}_\text{BS}
 ) \log(\mathcal{N}_\text{MS}
  \mathcal{N}_\text{BS}
 ),
\]
where \(\zeta > 0\) is a universal constant, and \(\zeta = \sqrt{\frac{\mathcal{N}_\text{MS}\mathcal{N}_\text{BS}
 }{|\mathbf\Omega|}}\).

The proof leverages fundamental results from low-rank matrix completion theory. Specifically, the nuclear norm \(\|\widetilde{\textbf{Y}}\|_*\) serves as a convex surrogate of the non-convex rank function. Under a uniform random sampling model, the projection operator \(\mathcal{P}_{\mathbf\Omega}\) satisfies the RIP for rank-\(\widehat{\mathcal{R}}_t\) matrices with high probability, provided that the number of observed entries \(|\mathbf\Omega|\) is sufficiently large. Consequently, the optimization problem in (\textcolor{blue}{11}) can be reformulated as a constrained nuclear-norm minimization problem. By imposing the standard incoherence condition on \(\widetilde{\textbf{Y}}_t\), that is, its left and right singular vectors are not excessively aligned with the canonical basis—exact econstruction is guaranteed with the derived sample complexity. is chosen to balance the trade-off between data fidelity and low-rank regularization.
This result demonstrates that the proposed framework achieves reliable reconstruction with only 
 \(O(\widehat{\mathcal{R}}_t (\mathcal{N}_\text{MS}+ \mathcal{N}_\text{BS}) \log(\mathcal{N}_\text{MS}\mathcal{N}_\text{BS}))\) measurements, which is substantially lower than the \(O(\mathcal{N}_\text{MS}\mathcal{N}_\text{BS})\) samples required under full observation.
 
Such reduced sampling complexity highlights the scalability and efficiency of the proposed method, particularly for large-scale mmWave channel estimation, where exhaustive sampling is computationally prohibitive.
\subsection{Complexity Ananysis}  
The proposed rank-aware framework attains high computational efficiency by employing dynamic rank estimation (\(\widehat{\mathcal{R}}_t\)) and tightly coupled processing stages. The complexity of each part is analyzed as follows.  

\textit{Rank Estimation and Matrix Completion}: This phase involves two main components: rank prediction and matrix reconstruction.
First, the autoregressive (AR) model-based rank prediction requires only \(\mathcal{O}(j)\) operations where \(j \leq 3\)  is the AR order. This cost is negligible compared to the subsequent steps. Second, solving the rank-one decomposition via the ADMM algorithm entails two dominant operations: partial SVD with complexity \(\mathcal{O}(\mathcal{L}_1\mathcal{L}_2\widehat{\mathcal{R}}_t)\)), and block coordinate descent (BCD) updates with complexity \(\mathcal{O}(\widehat{\mathcal{R}}_t\mathcal{L}_1\mathcal{L}_2)\)). Here, \(\mathcal{L}_1\) and \(\mathcal{L}_2\) denote the dimensions of the angle-domain overcomplete dictionary, satisfying (\(\mathcal{L}_1 \ll \mathcal{N}_{\text{BS}}\) and \(\mathcal{L}_2 \ll \mathcal{N}_{\text{MS}}\)). Empirical convergence is typically achieved within \(T\) iterations, resulting in an overall complexity of \(\mathcal{O}(T \widehat{\mathcal{R}}_t \mathcal{L}_1\mathcal{L}_2)\). Compared with conventional low-rank matrix completion (LRMC) algorithms, the AR-based initialization substantially accelerates convergence and reduces the total iterations. 

\textit{Rank-Aware Sparse Recovery}: Leveraging the established equivalence between matrix rank and sparsity, i.e., (\(\text{rank}(\mathbf{H}_t) = \|\widetilde{\mathbf{H}}_t\|_0 = \widehat{\mathcal{R}}_t\) (see Sec. II-B), the proposed RA-BOMP algorithm adopts \(\widehat{\mathcal{R}}_t\) as an exact stopping criterion. Each iteration consists of correlation computation with complexity (\(\mathcal{O}(\mathcal{L}_1\mathcal{L}_2\mathcal{N}_{\text{BS}})\)) and least-squares (LS) updates with complexity (\(\mathcal{O}(\mathcal N^2_{\text{set}}\mathcal{N}_{\text{BS}} +\mathcal N^3_{\text{set}})\)), where \(\mathcal N_{\text{set}}\) denotes the current size of support set. Consequently, the total complexity is expressed as \(\mathcal{O}( \widehat{\mathcal{R}}_t^3 + \widehat{\mathcal{R}}_t \mathcal{L}_1\mathcal{L}_2\mathcal{N}_{\text{BS}} )\). Unlike conventional OMP algorithms that rely on an overestimated sparsity upper bound, the proposed rank-aware design achieves reduced complexity and improved computational determinism.  

\textit{Spatial Mode Matching and Dictionary Reduction}:
The RAMMD module further decreases computational burden by concentrating on dominant angular clusters. Energy-based sorting is first applied to identify the top \(\widehat{\mathcal{R}}_t\) clusters, which incurs a cost of (\(\mathcal{O}(\mathcal{L}_1 \log \mathcal{L}_1)\)). Then, beamforming matrices are reconstructed only within the localized angular regions \([\hat{\theta}_{\tau}^{\text{min}}, \hat{\theta}_{\tau}^{\text{max}}]\), resulting in complexity\(\mathcal{O}(\widehat{\mathcal{R}}_t \mathcal{L}_1 \mathcal{N}_{\text{BS}}\mathcal{N}_{\text{MS}})\). By adaptively restricting the search to \(\widehat{\mathcal{R}}_t\) dynamic subregions, the proposed method effectively eliminates the exhaustive full-angle scanning overhead present in traditional designs.

\section{NUMERICAL RESULTS}
\subsection{Simulation Settings}
In this section, the performance of the proposed scheme is evaluated through numerical simulations. We consider a hybrid architecture for an mmWave MIMO system. The base station (BS) is equipped with ${\mathcal{N}_\text{BS}}\in \{64,32,16,8,4\}$ antennas and ${\mathcal{M}_\text{BS}}\in \{8,4\}$ RF chains, while MS has ${\mathcal{N}_\text{MS}}\in\{32,16,8,4\}$ antennas and $\{8,4\}$ RF chains. The system operates at a center frequency of 28 GHz with a sampling period of 0.1$\mu s$. 

To assess estimation accuracy, training time instances are distributed across both complex and simple propagation scenarios, allowing us to examine the impact of rank variations on performance. Channel gains follow a complex Gaussian distribution. Departure and arrival angles are uniformly drawn from $(-\pi /2,\pi /2)$, and additive noise is modeled as independent complex Gaussian white noise. To evaluate the performance of matrix construction, the percentages of the missed/corrupt measurement data under dynamic conditions are set to be 10\%, 15\%, and 20\%, respectively. Accordingly, the missed/corrupted at static condition is uniformly set to be 5\%.

The normalized mean square error (NMSE) is adopted to quantify channel estimation performance, defined as:
\begin{equation}
 \mathrm{NMSE}={\text E}\left [\frac{||\textbf{H}_t-\hat{\textbf{H}}_t||_F^2}{||\textbf{H}_t||_F^2}\right ], 
 \end{equation}
 where ${{\textbf{H}}_t}$ denotes the true channel matrix and ${{\hat{\textbf{H}}_t}}$ is its estimate.
\subsection{Performance Analysis }
Fig. \textcolor{blue}{3} illustrates the probability of successful reconstruction of various channel estimation methods versus SNR in a high-mobility mmWave scenario. The proposed rank-aware scheme consistently achieves higher recovery rates across the entire SNR range compared with conventional methods, such as SPC-TDCS \cite{b37}, which rely solely on delay-domain sparsity and neglect spatial structure.
This performance gain stems from the joint exploitation of low-rank matrix completion and rank-aware sparse recovery, which fully leverages the spatio-temporal structure of mmWave channels. Even under rapid channel variations induced by mobility, the proposed approach maintains stable recovery performance by dynamically adapting the estimation strategy in response to the temporal evolution of channel rank. In contrast, existing algorithms that assume fixed sparsity or rank levels fail to capture such dynamics, resulting in degraded reconstruction probability, particularly in low-to-moderate SNR regimes.
Furthermore, the performance gap at low SNR highlights the critical role of the R1MC-based matrix completion phase, which provides reliable initialization by recovering the dominant channel subspace from severely incomplete and noisy observations. These results validate the effectiveness of incorporating rank-awareness and structured sparsity into mmWave channel estimation.

\begin{figure}[t]
	\centering 
	\includegraphics[width=8.5cm]{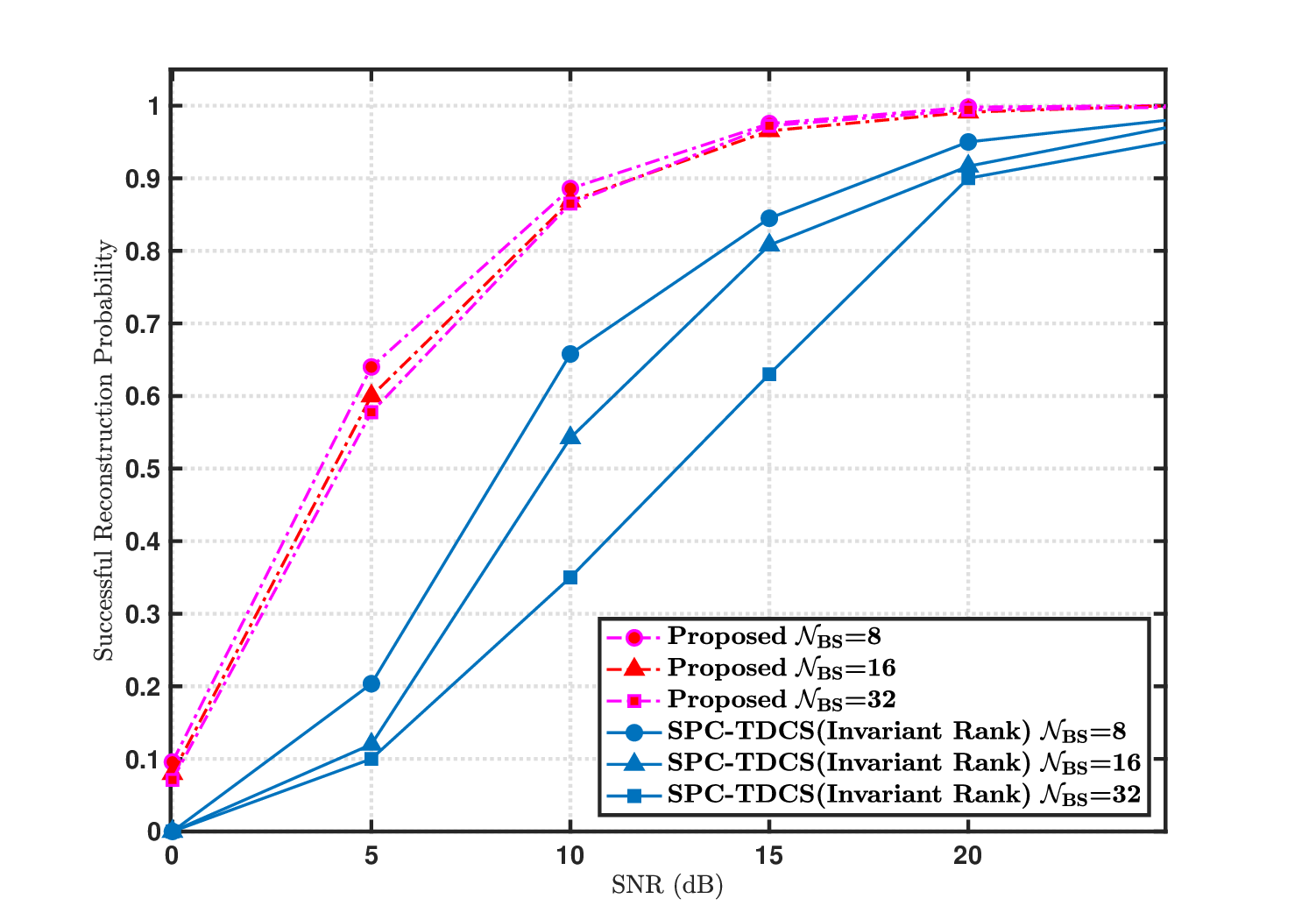}
	\caption{Successful reconstruction probability vs SNR with $v=120$ km/h, ${\mathcal{N}_\text{BS}}=8$, ${\mathcal{N}_\text{MS}}=8$. The 10\% measurement data are set to be missed/corrupted.}

\end{figure}

Fig. \textcolor{blue}{4} depicts the NMSE of channel estimation versus pilot overhead for the proposed scheme and benchmark algorithms. As observed, the NMSE of all schemes decreases with increasing pilot overhead. Notably, the NMSE of the proposed scheme stabilizes once the pilot overhead exceeds 6\%, indicating that only 6\% of this overhead is sufficient to achieve accurate channel estimation. This low pilot overhead requirement is primarily attributed to the combined two-phase channel estimation and the RAMMD module. These techniques enable better CSI estimation by selecting an appropriate number of pilot symbols from a limited subset of the angular domain. In contrast, the SPC-TDCS and Deep CNN-based methods \cite{b25} require at least 10\% pilot overhead, corresponding to a smaller number of pilot symbols, to reach comparable accuracy. While deep learning-based approaches can further improve estimation performance given sufficient pilot symbols, the SPC-TDCS method remains rank-invariant and does not exploit temporal rank dynamics.
\begin{figure}[t]
	\centering 
	\includegraphics[width=8.5cm]{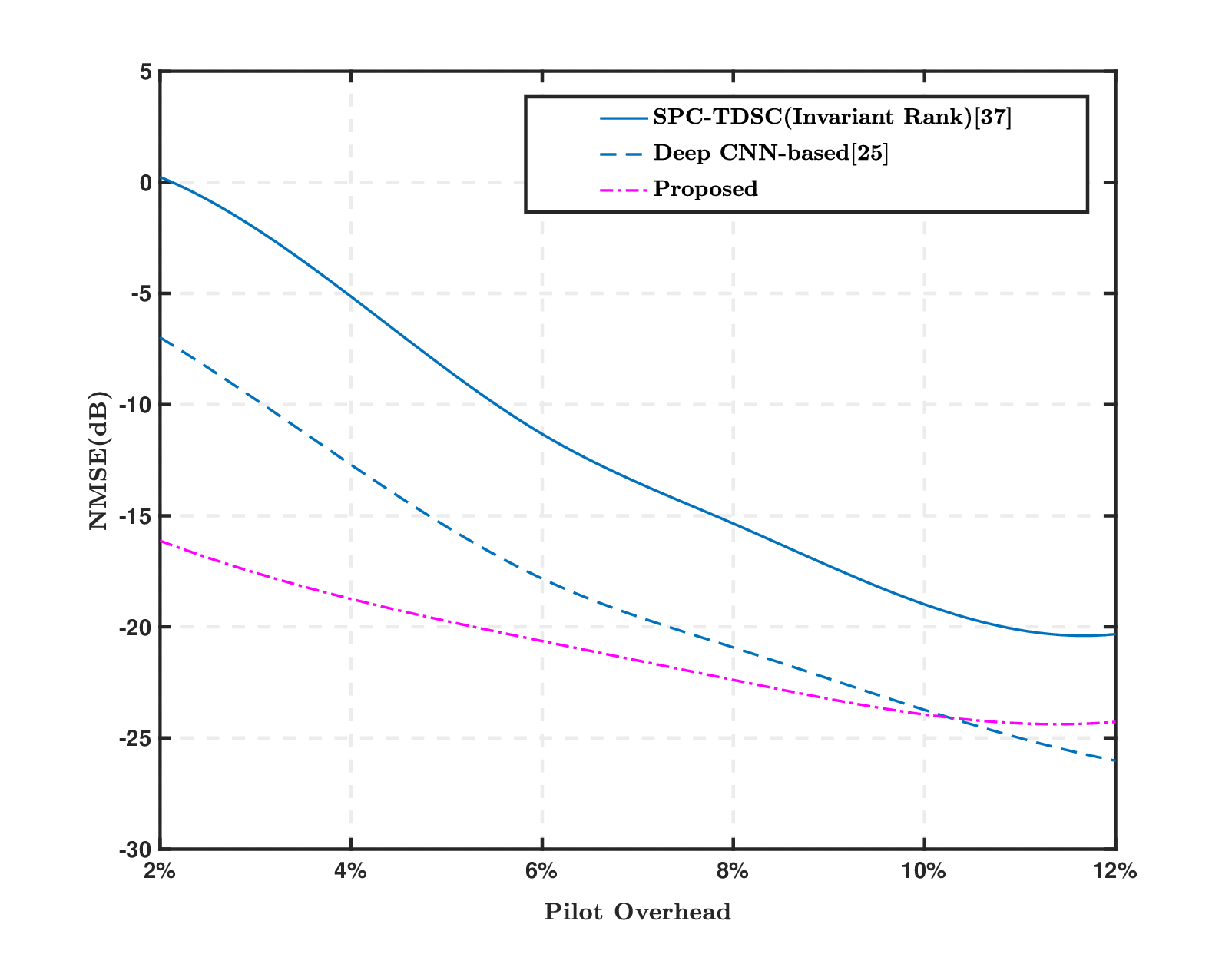}
	\caption{NMSE vs pilot overhead with $v=120$ km/h, SNR = 25 dB, ${\mathcal{N}_\text{BS}}=8$, ${\mathcal{N}_\text{MS}}=8$. The 15\% measurement data are set to be missed/corrupted.}
	
\end{figure}

Fig. \textcolor{blue}{5} compares the NMSE performance of the proposed scheme with OMP and SPC-TDCS for $\mathcal{N}_\text{BS} = 8$ and $\mathcal{N}_\text{BS} = 64$. Solid lines correspond to $\mathcal{N}_\text{BS} = 64$ and dotted lines to $\mathcal{N}_\text{BS} = 8$. Across different scenarios, simpler configurations achieve both lower latency and satisfactory accuracy, while larger antenna arrays yield significantly higher estimation precision. The performance advantage of SPC-TDCS arises from its effective utilization of structural channel characteristics in the measurement matrix design. By contrast, the proposed method further exploits temporal channel correlation rather than assuming a quasi-static channel, highlighting the benefit of capturing environmental dynamics and the necessity of re-estimating channel angles through rank-adaptive processing.
\begin{figure}[t]
	\centering 
	\includegraphics[width=8.5cm]{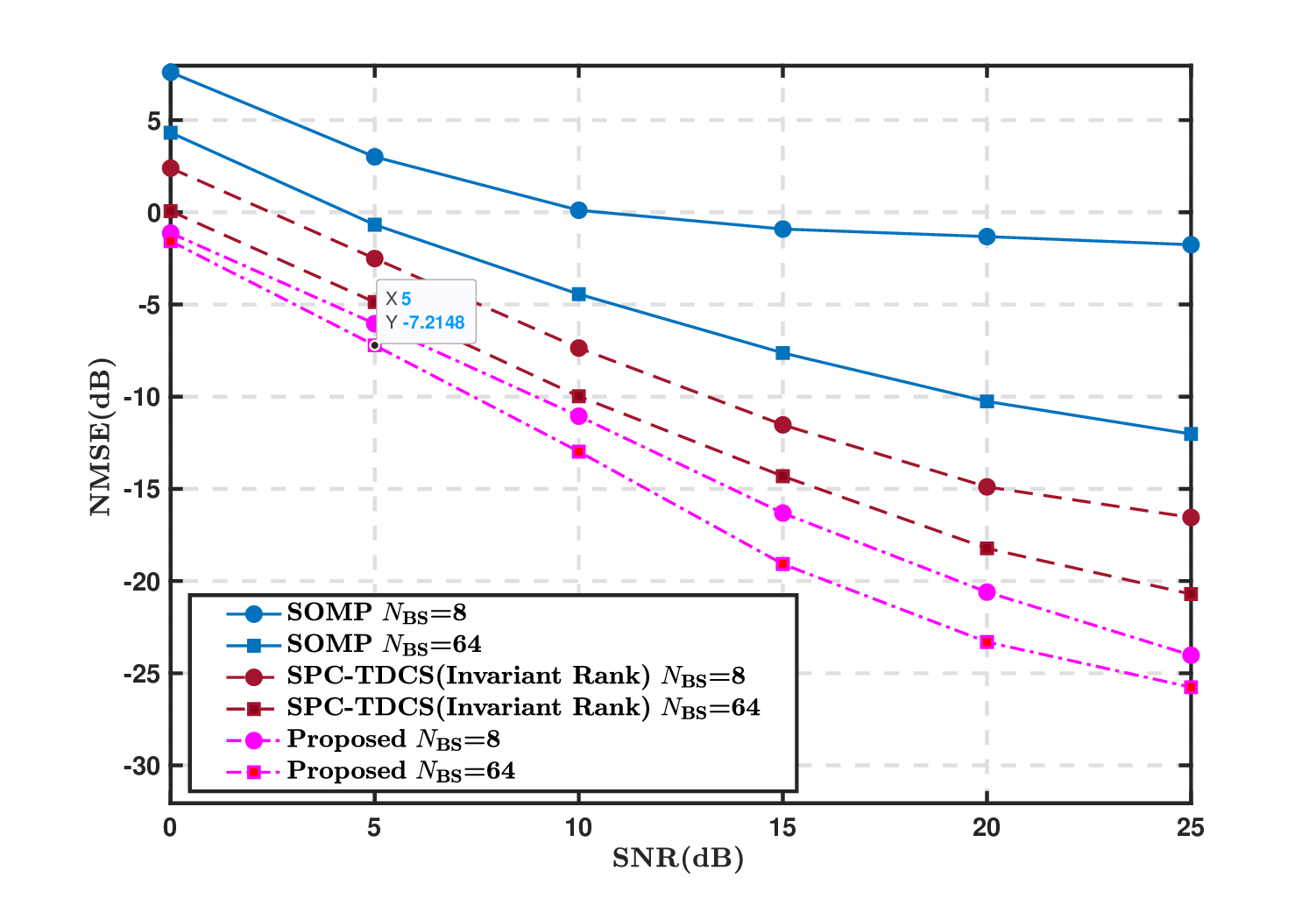}
	\caption{NMSE performance vs SNR with $v=120$ km/h, $\mathcal{N}_\text{MS}=8$, ${\mathcal{N}_\text{BS}}=8$ and $\mathcal{N}_\text{BS}=64$. The 20\% measurement data are set to be missed/corrupted.}

\end{figure}

\begin{figure}[t]
	\centering 
	\includegraphics[width=8.5cm]{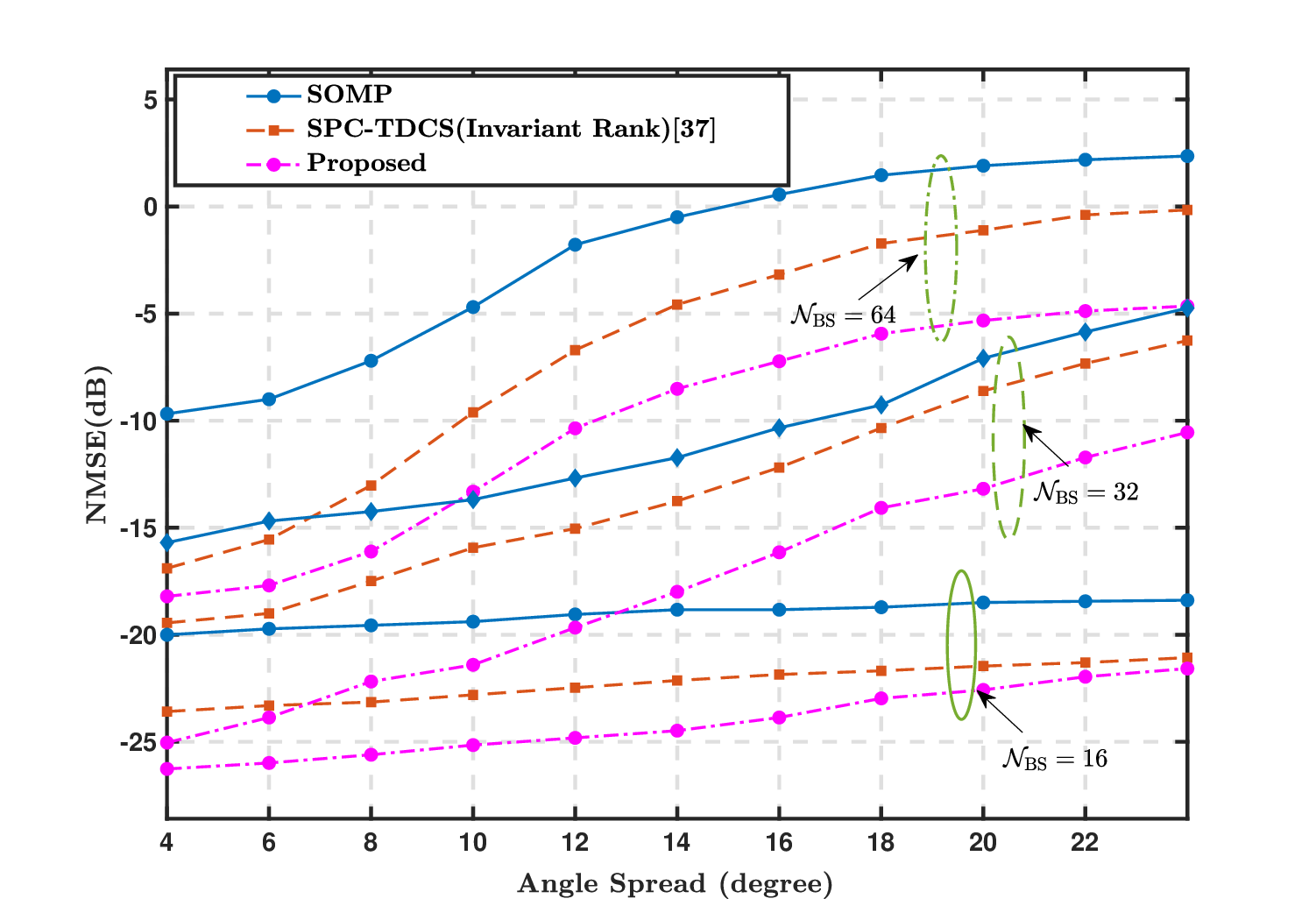}
	\caption{NMSE performance vs the angle spread with $v=120$ km/h, ${\mathcal{N}_\text{MS}}=8$. The 10\% measurement data are set to be missed/corrupted.}
	
\end{figure}
Fig. \textcolor{blue}{6} illustrates NMSE performance as a function of angle spread $\Delta\theta$ for different antenna configurations, specifically $\mathcal{N}_\text{BS}\in\{16, 32, 64\}$. It is observed that wider angle spreads lead to higher NMSE, due to the degradation of spatial correlation, resulting in larger estimation errors. Across all examined $\Delta\theta$ ranges and antenna configurations, the proposed algorithm consistently outperforms the conventional SOMP method. Moreover, as the number of antennas increases, the performance gap between our proposed scheme and other methods becomes more pronounced. This improvement is attributed to more accurate rank estimation within the proposed framework, which enhanced the correlation exploitation of matrix ranks, confirming the significant potential of our approach for practical massive MIMO deployments.
\begin{figure}[t]
	\centering 
	\includegraphics[width=8.5cm]{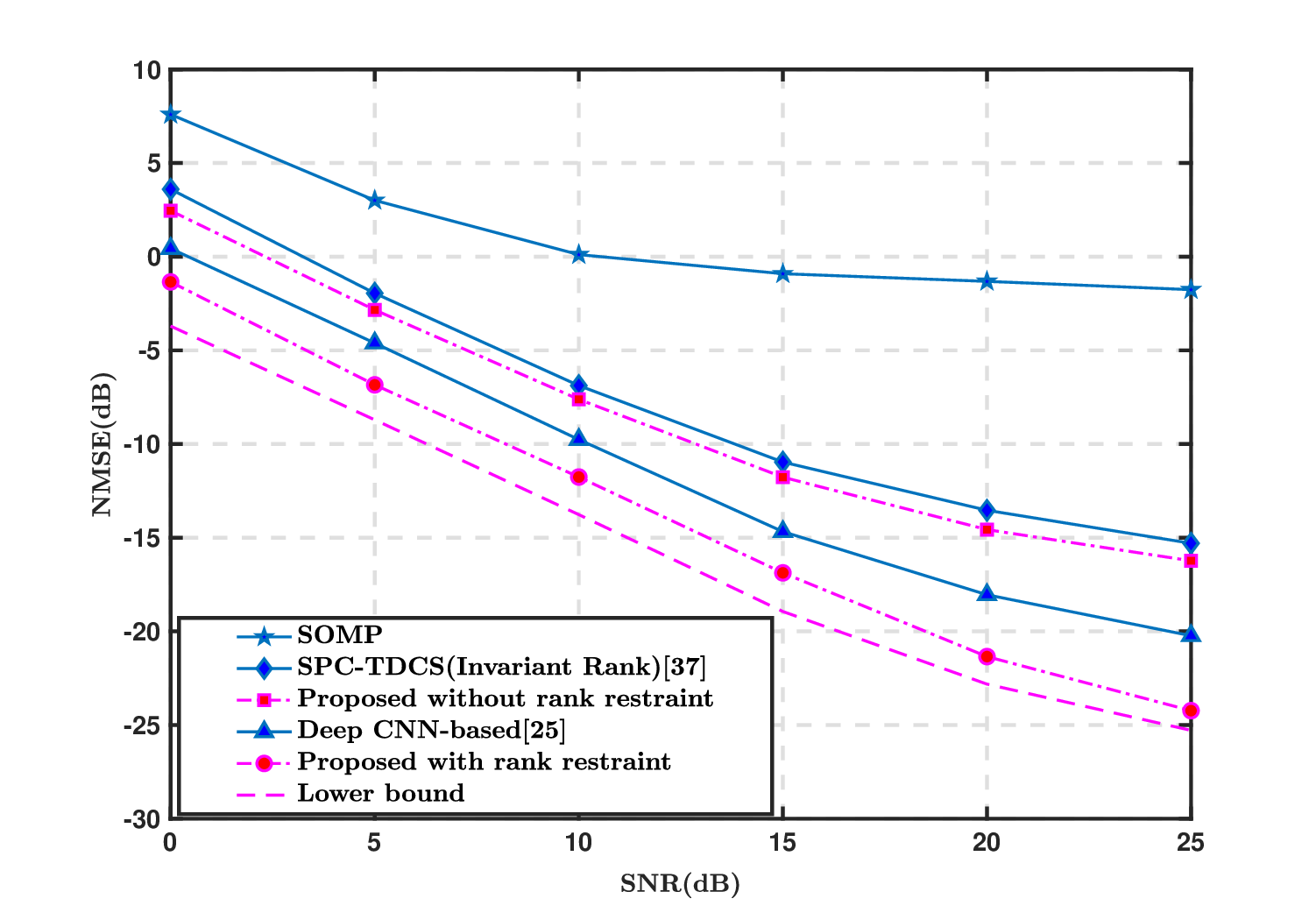}
	\caption{NMSE performance vs SNR with $v=120$ km/h, ${\mathcal{N}_\text{BS}}=8$, ${\mathcal{N}_\text{MS}}=8$. The 10\% measurement data are set to be missed/corrupted.}

\end{figure}

To further validate the effectiveness of the proposed rank-restraint mechanism, we conducted an ablation experiment by disabling rank restraint. In this configuration, the rank information is not propagated across successive algorithms; equivalently, RA-BOMP and RAMMD are configured to converge to arbitrary values without enforcing rank-based termination. This setting assumes that the angular transformation is always slower than the gain variation, thus neglecting the temporal rank evolution.

As illustrated in Fig. \textcolor{blue}{7}, the estimation accuracy of SOMP and SPC-TDCS deteriorates rapidly with increasing user mobility, supporting the importance of rank adaptation. In contrast, integrating a rank-aware estimator within the proposed framework leads to a noticeable improvement in NMSE performance. This improvement becomes more pronounced at higher SNRs, as the more effective use of channel parameters enables more accurate tracking of the underlying spatio-temporal variations.

\begin{figure}[t]
	\centering 
	\includegraphics[width=8.5cm]{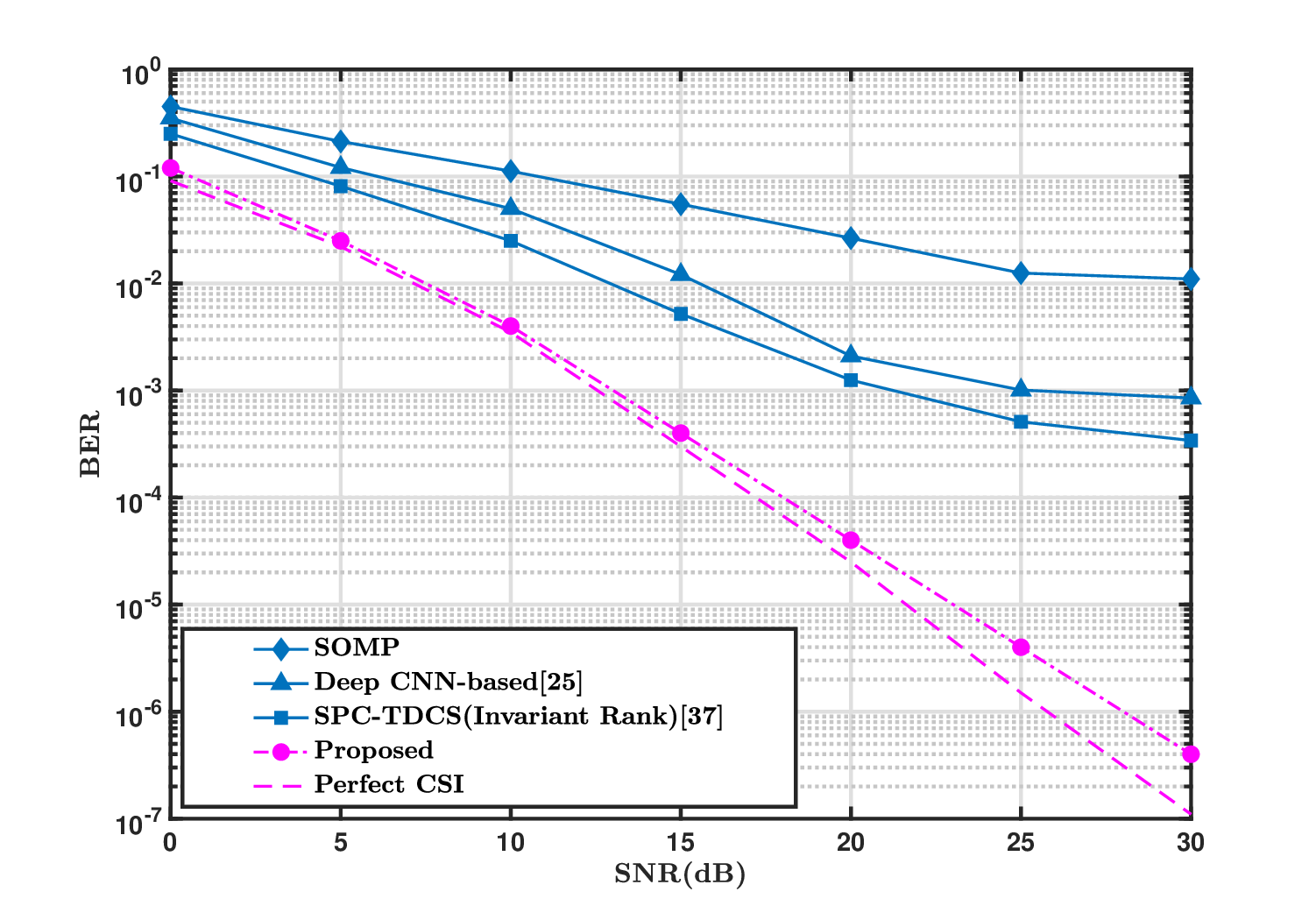}
	\caption{BER performance vs SNR, $v=120$ km/h, ${\mathcal{N}_\text{BS}}=8$, ${\mathcal{N}_\text{MS}}=8$. The 10\% measurement data are set to be missed/corrupted.}
	
\end{figure}

Fig. \textcolor{blue}{8} illustrates the bit error rate (BER) performance comparison among different channel estimation schemes. The proposed method achieves the lowest BER across all SNR levels, approaching the performance bound of perfect CSI. This gain arises from the accurate channel reconstruction achieved through the joint operation of the R1MC and RA-BOMP modules, which effectively preserve essential channel structures under high-mobility conditions. In contrast, SOMP and SPC-TDCS exhibit high BER floors due to their inability to accommodate rank variations or spatial mismatches, where the CNN-based deep learning method suffers from limited generalization capability, particularly at high SNR. Notably, the BER of the SOMP scheme remains above $10^{-2}$ even in high-SNR regimes, indicating its difficulty in maintaining reliable communication under practical conditions. These results collectively highlight the robustness, adaptability, and reliability of the proposed rank-aware estimator in realistic high-mobility mmWave MIMO environments.

\subsection{Performance Evaluation for Algorithm Parameters}
\begin{figure}[t]
	\centering 
	\includegraphics[width=8.5cm]{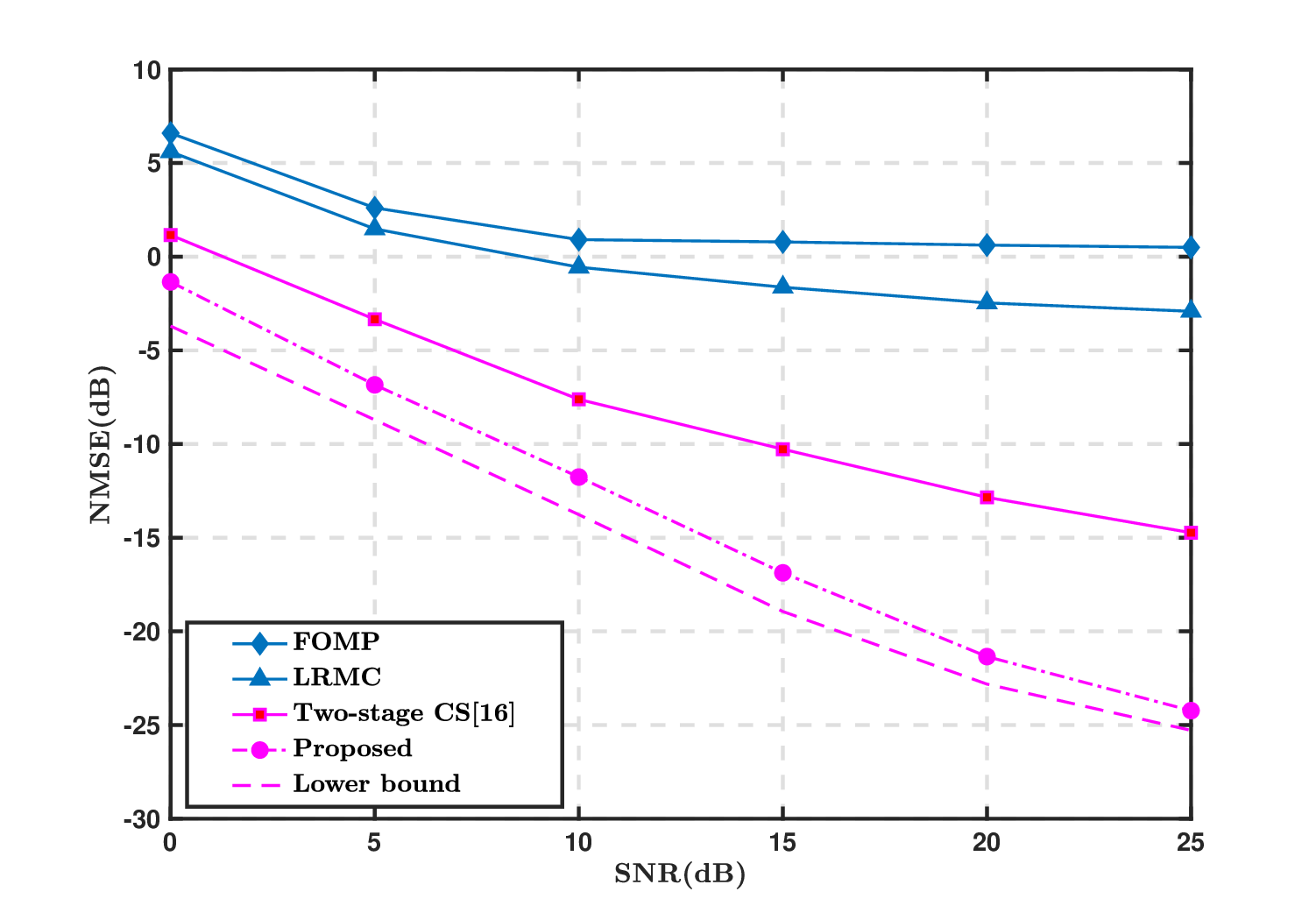}
	\caption{NMSE performance vs SNR in a static channel, ${\mathcal{N}_\text{BS}}=8$, ${\mathcal{N}_\text{MS}}=8$. The 5\% measurement data are set to be missed/corrupted.}
\end{figure}

In Fig. \textcolor{blue}{9}, the NMSE performance of various channel estimation algorithms is shown for a static mmWave channel scenario, where the channel rank remains constant and no temporal variation or dynamic rank adaptation is present. Even in this simplified condition, the proposed method substantially outperforms conventional techniques such as FOMP, LRMC, and the two-stage compressed sensing (CS) approach in \cite{b16}. The superior performance can be attributed to the joint use of R1MC and RA-BOMP, which effectively leverage the known rank information to impose tighter constraints on both the matrix completion and sparse recovery stages. In contrast, LRMC and \cite{b16} either neglect rank information or adopt heuristic sparsity levels, resulting in suboptimal reconstruction accuracy. By integrating rank priors as hard constraints, the proposed approach achieves more precise reconstruction and reduced estimation error. As the SNR increases, the NMSE curve of the proposed method approaches the theoretical lower bound, validating its ability to accurately extract the channel structure even in the absence of temporal dynamics or rank evolution.

\begin{figure}[t]
	\centering 
	\includegraphics[width=8.5cm]{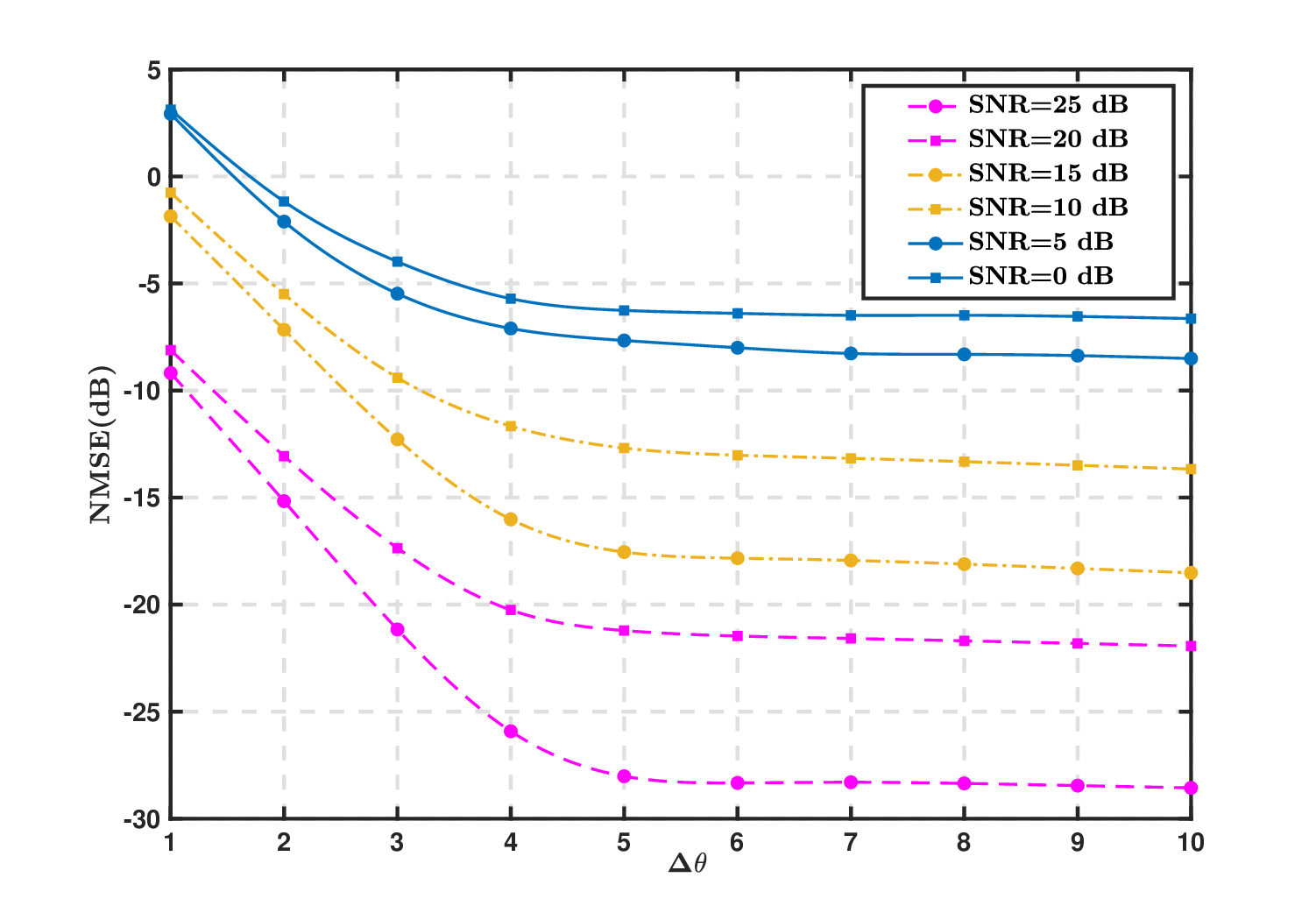}
	\caption{NMSE performance vs $\Delta\theta$ with $v=120$ km/h, ${\mathcal{N}_\text{BS}}=8$, ${\mathcal{N}_\text{MS}}=8$.}
	
\end{figure}
Fig. \textcolor{blue}{10} investigates the influence of the angular spread $\Delta\theta$ on the performance of the RAMMD algorithm, varying from $1^{\circ}$ to $10^{\circ}$ at a speed of 120 km/h. The results clearly indicate that $\Delta\theta$ has a pronounced impact on channel estimation accuracy. Specifically, increasing $\Delta\theta$ initially enhances estimation performance, as a moderate angular expansion captures additional beam components outside the initially constrained region. The NMSE decreases rapidly as $\Delta\theta$ increases from $1^{\circ}$ to $5^{\circ}$ , and then exhibits a slower decline for $\Delta\theta>5^{\circ}$. Accordingly, $\Delta\theta=5^{\circ}$ is adopted as a practical trade-off between estimation accuracy and computational complexity. Larger $\Delta\theta$ values require finer angular grids and thus higher computational cost; therefore, the chosen configuration achieves satisfactory estimation accuracy with minimal complexity overhead.

\begin{figure}[t]
	\centering 
	\includegraphics[width=8.5cm]{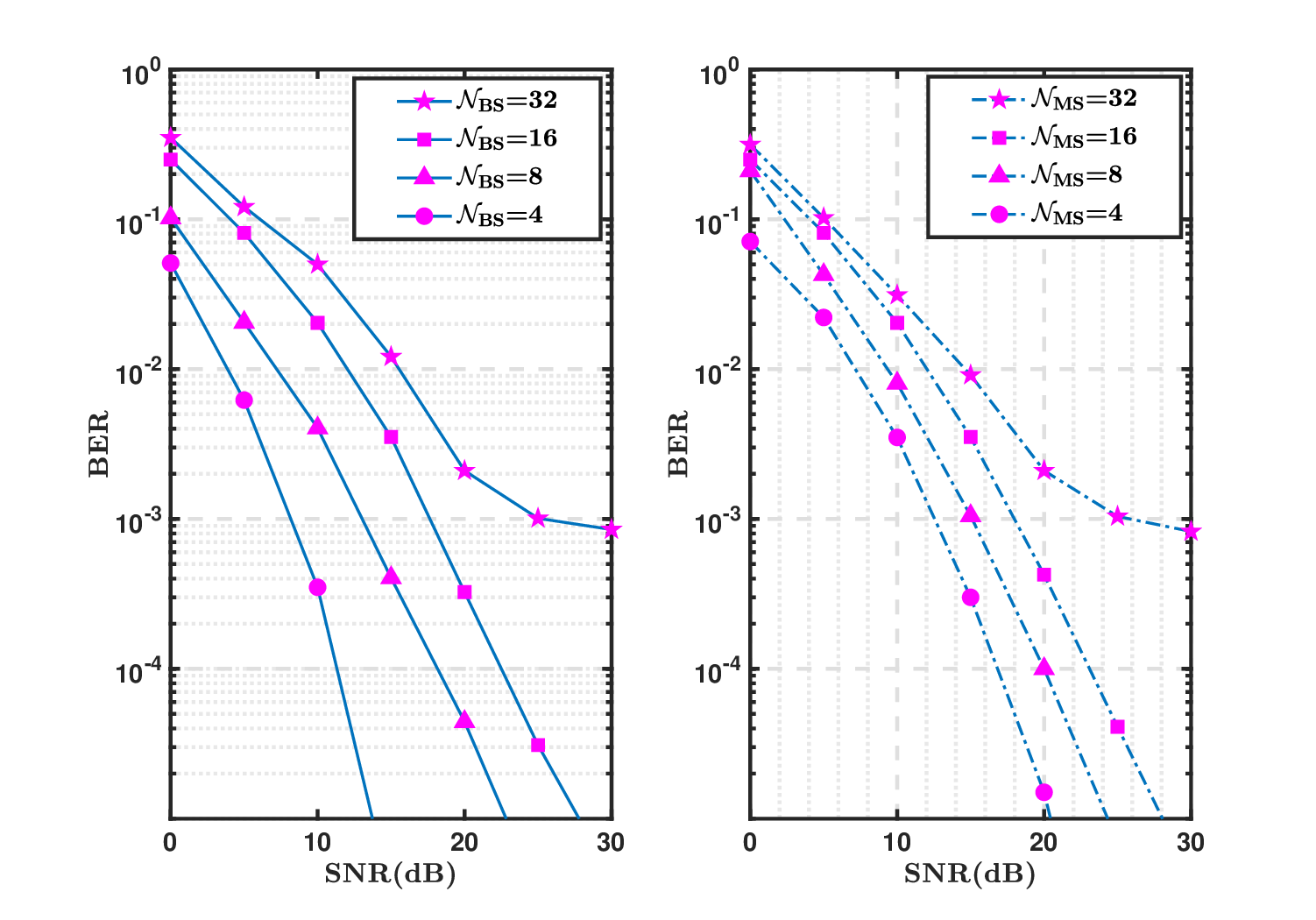}
	\caption{BER performance vs SNR at speed $v=120$ km/h. (a) $\mathcal{N}_\text{MS} =8$; (b) $\mathcal{N}_\text{BS}=16$}
	
\end{figure}

Fig. \textcolor{blue}{11} presents the BER performance of the proposed scheme under different antenna configurations. In Fig. \textcolor{blue}{11}(a), the number of receive antennas is fixed at $\mathcal{N}_\text{MS} = 8$, while the number of transmit antennas varies as $\mathcal{N}_\text{BS} = 4, 8, 16, 32$. The BER performance decreases as the number of transmit antennas increases. This behavior occurs because increasing $\mathcal{N}_\text{BS}$ effectively reduces the relative diversity gain available at the receiver. In Fig. \textcolor{blue}{11}(b), the number of transmit antennas is fixed at $\mathcal{N}_\text{BS} = 16$, and the number of receive antennas increases from 4 to 32. As shown in Fig. \textcolor{blue}{11}(b), the BER performance improves with larger $\mathcal{N}_\text{MS}$, indicating that increasing the number of receive antennas effectively enhances diversity gain and improves detection reliability. These results confirm that the proposed framework maintains robustness across a wide range of antenna configurations, providing a favorable balance between estimation accuracy and system scalability.

\section{CONCLUSION}

This paper presents a novel rank-aware framework for channel estimation and reconstruction in time-varying mmWave MIMO systems. The proposed approach decomposes the estimation process into two stages, namely observation matrix completion and sparse recovery. A robust rank-one matrix completion (R1MC) method is introduced to reconstruct the observation matrix. This facilitates efficient channel estimation and reconstruction with reduced computational complexity, achieved through the proposed framework and the rank-aware BOMP (RA-BOMP) algorithm. Furthermore, rank-awareness is integrated throughout the estimation process, leveraging the inherent low-rank and sparse structures of mmWave channels to improve both estimation stability and accuracy. An adaptive measurement matrix design based on rank prior further improves the estimation accuracy of the number of channel clusters. Compared to existing works, our analysis shows that the proposed scheme requires significantly fewer pilot measurements and outperforms conventional compressed sensing methods based on convex relaxation algorithms that rely solely on the channel sparsity. 

\end{document}